\documentclass[11pt]{article}

\usepackage[utf8]{inputenc}     
\usepackage[T1]{fontenc}        
\usepackage{hyperref}           
\usepackage{url}                
\usepackage{booktabs}           
\usepackage{amsfonts}           
\usepackage{nicefrac}           
\usepackage{microtype}          
\usepackage{lipsum}             
\usepackage{fancyhdr}           
\usepackage{graphicx}           
\graphicspath{{media/}}         
\usepackage{xcolor}
\usepackage{soul}
\usepackage{subcaption}
\usepackage{multirow}
\usepackage{multicol}
\usepackage{array}
\usepackage{tikz}
\usepackage{amssymb}
\usepackage{amsmath}
\usepackage{algorithm}
\usepackage{algpseudocode}
\usepackage{enumitem}
\usepackage{todonotes}
\usepackage{etoolbox}
\usepackage{longtable}
\usepackage{pdflscape}
\usepackage{colortbl}

\newcommand{\myrowcolour}{\rowcolor[gray]{0.925}}

\newtheorem{theorem}{Theorem}[section]

\newtheorem{remark}[theorem]{Remark}

\usepackage{geometry}
\hypersetup{
    colorlinks=true,
    linkcolor=blue,
    citecolor=blue,
    urlcolor=blue,
    pdftitle={Your Title},
    pdfauthor={Your Name, Coauthor One, Coauthor Two, Coauthor Three}
}

\title{Random walk based snapshot clustering for detecting community dynamics in temporal networks}

\author{
  Filip Bla\v skovi\' c\thanks{Zuse Institute Berlin, Berlin, Germany. \texttt{blaskovic@zib.de}} \and
  Tim O.F. Conrad\thanks{Zuse Institute Berlin, Berlin, Germany. \texttt{conrad@zib.de}} \and
  Stefan Klus\thanks{Heriot--Watt University, Edinburgh, UK. \texttt{S.Klus@hw.ac.uk}} \and
  Nata\v sa Djurdjevac Conrad\thanks{Zuse Institute Berlin, Berlin, Germany. \texttt{natasa.conrad@zib.de}}
}

\date{}

\begin{document}
\maketitle

\begin{abstract}
The evolution of many dynamical systems that describe relationships or interactions between objects can be effectively modeled by temporal networks, which are typically represented as a sequence of static network snapshots. In this paper, we introduce a novel random walk-based approach that can identify clusters of time-snapshots in which network community structures are stable. This allows us to detect significant structural shifts over time, such as the splitting or merging of communities or their births and deaths. We also provide a low-dimensional representation of entire snapshots, placing those with similar community structure close to each other in the feature space. To validate our approach, we develop an agent-based algorithm that generates synthetic datasets with the desired characteristic properties, enabling thorough testing and benchmarking. We further demonstrate the effectiveness and broad applicability of our technique by testing it on various social dynamics models and real-world datasets and comparing its performance to several state-of-the-art algorithms. Our findings highlight the strength of our approach to correctly capture and analyze the dynamics of complex systems.
\end{abstract}


\section{Introduction}

\emph{Temporal networks}, also known as \emph{time-evolving} or \emph{dynamic} networks, can effectively model various complex systems in which entities (represented as nodes or vertices) and the relationships between them (modeled by edges or links) change over time. In contrast to \emph{static networks}, which are useful for representing systems where the relationships are stable, such as electrical or signaling networks \cite{Boccaletti_2006, Lambiotte_2022}, temporal networks take into account the effects of time and offer a more realistic representation of complex systems \cite{Holme_2012}. In social networks, for instance, the frequency of interactions between individuals may vary across different time periods: for example, colleagues may communicate frequently during work hours, while during weekends or vacations these contacts may decrease. This makes temporal networks an ideal approach for modeling time-dependent behavior. Temporal networks have many applications, including neural and brain networks \cite{Bassett_2011, Dimitriadis_2010}, biological ecosystems \cite{Bajardi_2011}, epidemic spreading \cite{Humphries_2021, Sanhedrai_2022}, and economic scenarios \cite{Kirman_1997}.

\begin{figure}[!ht]
  \centering
  \includegraphics[width=0.9\textwidth]{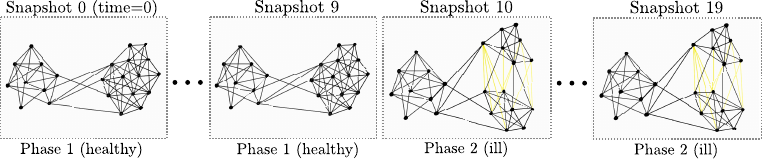}
  \caption{A temporal network with two phases. At the $10$th time step, the right community splits into two, marking the transition between the phases. The yellow edges in snapshots $10$ to $19$ represent edges that are not present anymore after the transition. Our new method can correctly identify the two phases and the time when the change occurred, and find a low-dimensional embedding of the snapshots.}
  \label{fig:intro_example}
\end{figure}

The analysis of network structures plays a crucial role in understanding the behavior of the underlying complex system. In particular, the identification of \emph{communities} (also called \emph{clusters} or \emph{modules}), which are defined as densely connected groups of nodes, reveals the systems' structural and functional organization \cite{Newman_2010}. Studying how these communities form, dissolve, and transform over time allows us to gain valuable insights into the underlying dynamics and interaction patterns \cite{Rossetti_2018}. For instance, the human microbiome is a complex microbial ecosystem whose composition and stability are closely linked to an individual’s health status\cite{Coyte_2015, Levy_2013, Faust_2012}. Network-based representations of microbial interactions enable researchers to investigate how diseases\cite{Baldassano_2016, Lam_2022, SanchezAlcoholado_2020, Steinway_2015} or medical treatments \cite{Shaw_2019, Steinway_2015} affect these communities and induce structural changes. In network terms, during healthy periods, microbial groups interact intensively between their members and form communities in the network. However, when the individual becomes ill, these interactions within the existing communities decrease, while interactions among other microbial groups increase, leading to a new stable structural configuration in the network \cite{Hsiao_2014}. In Figure \ref{fig:intro_example}, we show a simplified illustration of this process, where we consider a temporal network to be a sequence of static networks over a discretized time interval $[0, 19]$. Initially, the network features two communities (snapshots $0$ to $9$) corresponding to the "healthy" state and as the network evolves, one community splits into two smaller ones, leading to a new stable structural configuration (snapshots $10$ to $19$ that describes an "ill" state.

Change point detection plays a critical role across a wide range of application domains. For instance, it is well known that epileptic seizures are preceded by characteristic changes in brain dynamics, which can be effectively modeled using network-theoretic approaches \cite{Tauste_2018, Rungratsameetaweemana_2022, Cribben_2016}. A deeper understanding of these tipping points and the associated transitions in brain activity could greatly enhance seizure prediction and, consequently, improve the health and quality of life for individuals affected by neurological disorders. In addition, financial systems often exhibit abrupt phase transitions, such as those observed in interbank connectivity networks, where small perturbations may either remain localized or, upon crossing a critical threshold, trigger widespread cascading failures \cite{Acemoglu_2015, Amini_2016}. Lastly, in environmental science, network-theoretic tools have been extensively applied to study cascading tipping points (domino effects) in climate systems, such as those induced by global warming, which may lead to severe and irreversible impacts on ecosystems \cite{Wunderling_2021, Kronke_2020}. The topology of environmental networks, as well as the structural changes they undergo, plays a crucial role in determining system resilience and the likelihood of such critical transitions.

Despite recent advances in the field, the identification of communities in temporal networks and tracking their evolution over time remains a challenging problem \cite{Rossetti_2018}, particularly when dealing with large-scale and noisy datasets stemming from real-world systems. To address this challenge, we propose a method that clusters time-snapshots into so-called \textit{phases}, during which network communities remain largely unchanged. Each phase represents a stable network configuration, while transitions between different phases capture important changes in the network's community structure. The identification of phases allows us to significantly reduce the computational cost of performing community detection in every single snapshot. Instead, for every phase, we analyze only one (representative) snapshot or, alternatively, a (weighted) aggregated network obtained from all other snapshots of that phase. In the community dynamics of the temporal network shown in Figure \ref{fig:intro_example}, we can distinguish between two phases: a "healthy" phase with two communities and an "ill" phase with three communities.

\begin{figure}[!ht]
  \centering
  \includegraphics[width=0.95\textwidth]{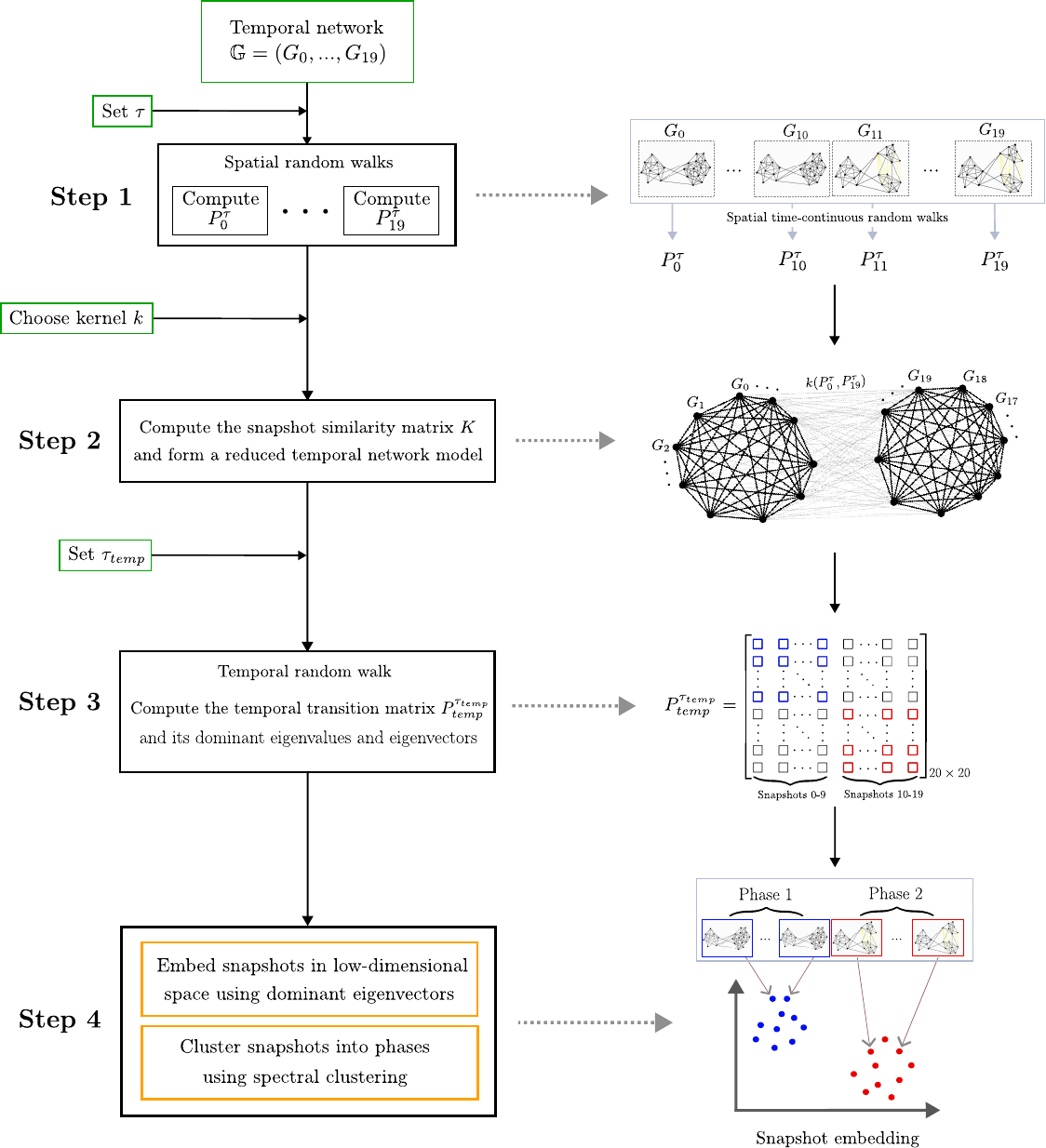}
  \caption{A schematic overview of our method demonstrated on a toy example consisting of 20 snapshots. First, we independently compute transition matrices of spatial random walks for each snapshot. We then perform pairwise comparisons between the snapshots using an appropriate similarity measure $k$. Finally, we construct a temporal transition matrix, apply spectral clustering to identify phases, and obtain low-dimensional representations of entire snapshots. Green blocks denote input data; orange blocks indicate final outputs.}
  \label{fig:method_illustration}
\end{figure}

In this work, we introduce a novel random walk-based approach that captures similarity between community structures of networks from different snapshots. We consider independent \textit{spatial random walks} on each of the snapshots to analyze their community structure and obtain transition matrices.  Our method uses the property that network communities are encoded in the block structure of the transition matrix. Thus, by comparing transition matrices of networks from different snapshots, we can detect changes in the communities over time. To quantify this, we introduce a measure of similarity between snapshots, such that snapshots with (almost) the same community structure have high similarity and snapshots with very different community structure have low similarity. We use these values to introduce a new static network, where nodes correspond to snapshots of the temporal network and we define edge strengths between these nodes using the snapshot similarity. This new network serves as a reduced model of the original temporal network and it encodes information about phases of the temporal network. That is, nodes corresponding to the snapshots of the same phase will be densely connected by edges with large weights and form a community within the reduced model, while snapshots with very different communities will be weakly connected. Finally, we define \textit{a temporal random walk process} on this new static network and use spectral clustering to identify its communities, i.e., phases of the original temporal network. In Figure \ref{fig:method_illustration}, we show a schematic illustration of our method, which consists of applications of the two distinct random walk processes mentioned above: a spatial random walk and a temporal random walk.

Additionally, our approach extends the concept of spatio-temporal exploration by introducing time-continuous random walks for both types of processes. This incorporates a time resolution parameter that can regulate the method's sensitivity to detect variations in the community structure and enable phase detection across multiple scales, from fine-grained to coarse-grained resolutions.
Similar concepts of spatio-temporal exploration for temporal clustering have been considered in previous work \cite{Gomez_2013, Sikorski_2021, Trower_2024, Froyland_2024}. However, the networks produced in these approaches can become exceptionally large when dealing with a high number of nodes and many snapshots, resulting in significant computational costs for their analysis. In contrast, our method ensures that the size of the reduced model remains independent of the node set size, making it more suitable for larger datasets, often associated with real-world systems. To further illustrate the efficacy of our method, we introduce a new benchmark generator for producing synthetic temporal network data using an agent-based approach. This generator is essential for testing our proposed method under controlled conditions. Synthetic datasets generated with this method form a robust foundation for evaluating our approach and comparing it with other methods.

\vspace{0.2cm}
\noindent The key contributions of our work are:
\vspace{0.2cm}

\begin{enumerate}[leftmargin=3.5ex, itemsep=0ex, topsep=0.5ex, label=\roman*)]
    \item We \textbf{introduce a new random walk-based method for identifying phases in temporal networks}, i.e., periods of stability in the community structure. Our method \textbf{improves robustness to noise and small fluctuations} and offers a more stable view of the underlying community dynamics by filtering out short-term variations. Our method not only detects these stable phases, but also \textbf{effectively captures significant structural transitions}, such as the splitting or merging of communities or their births and deaths, as well as meta-events, e.g., persistent communities or recurring community structures.
    
    \item Using our method, we can find a \textbf{low-dimensional embedding of network snapshots} that preserves the similarity between their underlying community structures, enabling effective comparison and analysis across time. Compared to existing approaches, our method reduces computational costs, making it more efficient for analyzing large-scale temporal networks. In particular, clustering snapshots into phases can serve as a way to \textbf{reduce the dimensionality of the temporal network data}, where instead of treating every snapshot individually, we can  compress the temporal sequence of networks into a smaller set of phases.
    
    \item We present a \textbf{novel benchmark generator} based on an agent-based model, that can simulate a wide range of dynamic community behaviors in temporal networks. Our benchmark generator is highly versatile and has the potential to be used as a robust tool for evaluating similar methods. This adaptability is crucial, as there is a growing need for standardized benchmarks to evaluate and compare different methods for analyzing temporal networks.
\end{enumerate}

Our paper is structured as follows: We first introduce our novel method for clustering network snapshots in Section~\ref{sec:methods}. Then, in Section \ref{sec:syntdataexp}, we present a short overview of existing benchmark algorithms, introduce a new benchmark generator, and validate our method using the synthetic networks obtained in this way. We continue the validation in Section \ref{sec:realdataexp} by applying our method to networks derived from social dynamics models and real-world datasets, showcasing its practical applicability and versatility across different domains. To situate our work within the existing body of research, Section \ref{sec:related_work_and_comparison} offers an overview of related work alongside a comparative analysis. We examine existing approaches, highlighting how our method not only addresses current gaps but also demonstrates superior performance, thereby underscoring its contributions to the field of dynamic network analysis. Finally, in Sections \ref{sec:discussion} and \ref{sec:conclusion} we present a brief discussion and conclusion, and suggest further research directions.

\section{A Novel Method for Clustering Time-Snapshots in Temporal Networks}
\label{sec:methods}

In this section, we present our new method \emph{Local Neighborhood Exploration (LNE)}  and its special case \emph{Invariant Measure Comparison (IMC)}. We illustrate its key steps and capabilities with a guiding example on a synthetic temporal network, demonstrating how it identifies transitions and clusters snapshots based on their community structures.

\subsection{Random Walks on Networks}
\label{sec:rw_on_networks}

We consider a temporal network $\mathbb{G}=(G_0,...,G_{M-1})$ to be a sequence of static networks $G_{\alpha}=(V_{\alpha}, E_{\alpha})$ at discrete times $\alpha\in\{0,...,M-1\}$, given by a set of nodes $V_{\alpha}$ and a set of edges $E_{\alpha}$. We assume that the set of nodes $V_\alpha=V$ is fixed during the entire evolution and that additionally networks in all snapshots are connected, aperiodic and undirected. Let $A_{\alpha}$ denote the (potentially weighted) adjacency matrix of a snapshot $G_{\alpha}$, where $A_{\alpha}(u,v)=w_{\alpha}(u,v)$ is the weight of the edge between nodes $u$ and $v$ (if the networks are unweighted, then $w_{\alpha}(u,v)=1$ if $(u,v)\in E_{\alpha}$ and $w_{\alpha}(u,v)=0$ otherwise). A standard discrete-time random walk process is defined by a Markov chain with the transition matrix $P_{\alpha} =\big[p_{\alpha}(u,v)\big]_{u,v \in V}$, given by
\begin{equation} \label{eq:RW}
    p_{\alpha}(u,v) = \frac{A_{\alpha}(u,v)}{d_{\alpha}(u)}, \quad \text{with } d_{\alpha}(u) = \sum_{v \in V} A_{\alpha}(u,v),
\end{equation}
where $d_{\alpha}(u)$ is the (weighted) degree of a node $u$ in a snapshot $\alpha$. This means that at each time
step the random walk process moves to one of the neighboring nodes that has been chosen randomly proportional to the edge weights. From a network perspective, communities are subsets of nodes, where nodes of the same community are densely connected to each other and sparsely connected to nodes in other communities. From a dynamic perspective, communities can be understood as metastable sets of the random walk process  \cite{Djurdjevac_2012, Sarich_2014, Klus_2024, Schuette_2013}, i.e., sets of nodes where the process remains ``trapped'' for a relatively long time before transitioning elsewhere. The community structure of the network is reflected in the long-term behavior of the random walk process.  More precisely, for metastable sets $A,B\subset V$, it holds that 
\begin{enumerate}
    \item[(a)] the residence probability in a metastable set $A$ is close to $1$, i.e., $ p_{\alpha}(A, A) = \mathbb{P}[X_T \in A \mid X_0 \in A] \approx 1$,
    \item[(b)] the transition probability between $A$ and $B$ is small, i.e., $p_{\alpha}(A,B) = \mathbb{P}[X_T\in B \mid X_0\in A]\ll 1$,
\end{enumerate}
for large enough $T$. The best full partition of the network $G_{\alpha}$ into $C_{\alpha}^1,...,C_{\alpha}^k$ disjoint communities, where  $\bigcup_{i=1}^k C_{\alpha}^i = V$ and $ C_{\alpha}^i \cap C_{\alpha}^j = \varnothing, \forall i \ne j $, is the one that maximizes the joint metastability of the sets
\begin{equation*}
    \mathcal{D}(C_{\alpha}^1, \dots, C_{\alpha}^k) = \sum\limits_{i=1}^k p_{\alpha}(C_{\alpha}^i, C_{\alpha}^i).
\end{equation*}
Since the networks we consider are connected and aperiodic, the random walk process under consideration is ergodic \cite{Djurdjevac_2012}. For ergodic processes, relations between a decomposition into metastable sets and spectral properties of the transition matrix $P$ have been established and used in many spectral clustering algorithms~\cite{vonLuxburg_2007}. In particular, the number of metastable sets (communities) can be inferred from the number of dominant (largest) eigenvalues of $P$ and decomposing dominant eigenvectors using their sign structure leads to the identification of these communities~\cite{vonLuxburg_2007, Sarich_2014}. The long-term behavior of the process is encoded in the left eigenvector corresponding to the unique, largest eigenvalue equal to $1$. This follows from the fundamental theorem of Markov chains, stating that every ergodic random walk process always converges to the unique \emph{invariant measure} (or stationary distribution) $\tilde{\mu}_{\alpha}$, where $\tilde{\mu}_{\alpha} P_{\alpha} = \tilde{\mu}_{\alpha}$, i.e., $\tilde{\mu}_{\alpha}$ is the left eigenvector corresponding to the largest eigenvalue of $P_{\alpha}$.

However, it is important to note that not all metastable sets of a random walk correspond to community structures within a network. Other configurations, such as long chains, can also trap the random walk process for extended periods, but do not represent true network communities \cite{Djurdjevac_2012}. To address this, we use time-continuous Markov processes (or Markov jump processes), as introduced in Sarich et al. \cite{Sarich_2014}, and for each network snapshot $G_{\alpha}$ at time $\alpha$, we introduce a time-continuous random walk process by defining a rate matrix
\begin{equation}
    L_{\alpha}(u,v)=
    \begin{cases}
        -\frac{1}{d_{\alpha}(u)}, & u=v, \\[1ex]
        \frac{A_{\alpha}(u,v)}{d_{\alpha}(u)^2}, & u\neq v, (u,v) \in E_{\alpha}, \\[0.5ex]
        0, & \text{otherwise.}
    \end{cases}
\label{eq:generator_matrix}
\end{equation}

Unlike standard discrete-time random walks (see Eq. \eqref{eq:RW}), this process has the advantage that it incorporates waiting times at each node. Specifically, the expected waiting time at a node $u$ is proportional to its degree $d_{\alpha}(u)$, allowing the process to move more quickly through loosely connected regions while slowing down and taking longer in densely interconnected structures, i.e., communities. This approach enhances the metastability of communities, thereby improving the performance of traditional spectral methods. For other examples of time-continuous random walk processes, we refer to Djurdjevac \cite{Djurdjevac_2012} and Sarich et al.\cite{Sarich_2014}

\subsection{Neighborhood Exploration using Random Walks: Spatial Random Walks}
\label{sec:spatial_rw}

The rate matrix $L_{\alpha}$ is also called the infinitesimal generator of a Markov process, since for a spatial exploration time $\tau >0$ it can generate the whole family of transition matrices
\begin{equation}\label{eq:P}
    P_{\alpha}^{\tau}=\exp(L_{\alpha}\tau).
\end{equation}
Efficient computation of the matrix exponential is possible using methods such as Krylov subspace techniques (since $L_{\alpha}$ is sparse in most applications), making the approach scalable even for large networks. The transition matrix $P_{\alpha}^{\tau}$ gives the probabilities of a process transitioning between two nodes after the time $\tau$, that is $P_{\alpha}^{\tau} = \big[p_{\alpha}^{\tau}(u,v)\big]$, where $p_{\alpha}^{\tau} (u,v)=\mathbb{P}[X_{\tau}=v \mid X_0=u]$. The choice of the time parameter $\tau$, which we refer to as \emph{spatial exploration time}, has a significant influence on the behavior of the process, and thus on the spectral properties of $P_{\alpha}^{\tau}$ and community identification. If $\tau$ is small, the random walk process explores only the immediate neighborhood and the transition probabilities reflect the local structure via short-range connections. If $\tau$ is large, on the other hand, the process has more time to explore the network and reflects its global long-term behavior such as metastability. Thus, the spectrum of $P_{\alpha}^{\tau}$ changes with the choice of the spatial exploration time and the optimal choice of $\tau$ depends on the desired resolution of the process. Of interest are large (but not too large) values of $\tau$ since these may reveal the underlying community structure. If there are $k$ communities (metastable sets) in the network, then the $k$ dominant (largest) eigenvalues $0 = \Lambda_1^{\alpha} > \Lambda_2^{\alpha} \geq \ldots \geq \Lambda_{k}^{\alpha}$ of $L_{\alpha}$ are close to zero and are separated by a spectral gap from the smaller eigenvalues. The implied timescales of this process are given by $1/|\Lambda_i^{\alpha}|$ for $i \geq 2$, and provide bounds for a good choice of $\tau$. Namely, the chosen resolution of the process reflects the granularity of a network \cite{Sarich_2014}, so that for finding the $k$ most metastable sets $\tau$ should be $1/|\Lambda_{k+1}^{\alpha}| < \tau < 1/|\Lambda_{k}^{\alpha}|$.

Since random walk processes explore the neighborhood of a node for a fixed $\tau$, which results in a constrained exploration range, we will call our method \emph{Local Neighborhood Exploration} (LNE). The LNE method uses transition matrices $P_{\alpha}^{\tau}$ as effective encodings for comparing community structure across snapshots. For a suitably chosen $\tau$, nodes that belong to communities exhibit a block structure in the transition matrix, potentially after an appropriate permutation of nodes. This comes as a consequence of the metastable behavior of the dynamics within communities: $\tau$ is chosen to be small enough so that the random walk does not leave the community due to its metastability, yet large enough to allow the process to sufficiently explore the host community. Thus, on a scale within the community, the probability distribution of the process's endpoint is approximately the same, regardless of the starting node. Specifically, let $G_0$ and $G_1$ be two snapshots, and $u,v$ two nodes belonging to the same community $C\subset V$ that does not change in both snapshots. Then, $P_0^{\tau}(u,:)\approx P_1^{\tau}(v,:)$, where $P(u,:)$ denotes the $u$th row of the matrix $P$. Hence, the communities accurately correspond to blocks in transition matrices and the stability in the community structures across snapshots is reflected in the similarity of the corresponding transition matrices. 

\subsection{Guiding Example}
\label{ex:guiding_example}
\begin{figure}[!ht]
  \centering
  \includegraphics[width=0.8\textwidth]{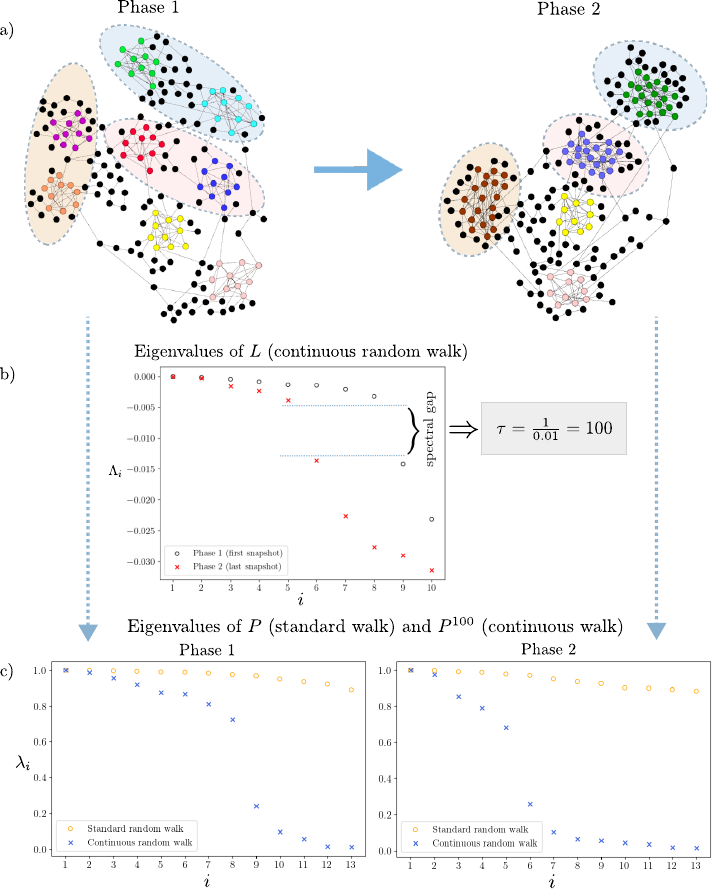}
  \caption{Illustration of the temporal network introduced in the guiding example (see Section \ref{ex:guiding_example}). a) Base networks and their communities for Phase 1 and Phase 2. Nodes colored in black do not belong to any communities. Other node colors indicate the community affiliation. Circled groups of nodes in Phase 1 indicate which communities will merge in the next phase. b) The dominant eigenvalues of the generator $L$ for the base network of Phase 1 (in black) and Phase 2 (in red). c) The dominant eigenvalues of the transition matrices for both phases, corresponding to the standard (orange circles) and time-continuous (blue crosses) random walks. Note that while eigenvalues are shown here for illustration, our proposed methods do not rely on their explicit computation.
  }
  \label{fig:guiding_example}
\end{figure}

As a guiding example, we consider a temporal network $\mathbb{G}=(G_0,...,G_{19})$ consisting of 200 nodes and 20 snapshots. The first ten snapshots are formed based on the static network ``Phase 1'' shown in Figure \ref{fig:guiding_example}. This network contains 8 distinct communities as highlighted in the figure with different community colors. Each snapshot from 0 to 9 is then obtained by introducing noise through the random addition and deletion of edges within communities in the base network, representing the first phase of the temporal network. Then, at time $\alpha=10$ new edges are introduced between selected pairs of communities, causing them to merge and resulting in the new base network, ``Phase 2'', with 5 communities. Snapshots 10 to 19 are now similarly formed by adding noise to this static network representing the second phase. Figure \ref{fig:guiding_example} shows the ten largest eigenvalues of the generator $L$ for the base network of Phase 1 and the base network of Phase 2, revealing spectral gaps after the 8th and 5th eigenvalue, respectively. To accurately detect network communities in both phases, we choose $\tau$ from the interval $1/0.014\approx 70\leq\tau\leq 250=1/0.004$, which is contained in both spectral gaps. In Figure \ref{fig:guiding_example} we show the 13 largest eigenvalues (blue crosses) corresponding to the time-continuous random walk of length $\tau=100$ for both phases and the 13 largest eigenvalues corresponding to the standard random walk (orange circles). We observe that the standard random walk, unlike the continuous approach, fails to capture the community structure, showing no clear spectral gaps, highlighting the higher generalizability of the continuous approach. Furthermore, it is important to note that adjusting the parameter $\tau$ modifies the resolution at which communities are detected. In order to keep a consistent resolution during the entire network evolution, we fix the granularity of our algorithm by choosing the same lag time $\tau$ for all snapshots.

\subsection{Exploring the Long-term Behavior of Random Walks}
\label{sec:IMC}

If the network exploration by the random walk process is performed for a sufficiently long period, then the probability of being at any node is approximately given by the invariant measure. As a special case of the LNE approach, obtained by letting $\tau$ tend to infinity, we consider a \emph{Invariant Measure Comparison (IMC)}, where we use the long-term probabilities given by the invariant measure to study network community dynamics. The invariant measure of a time-continuous random walk associated with an infinitesimal generator $L_{\alpha}$ is a probability distribution $\mu_{\alpha}$ such that $\mu_{\alpha}L_{\alpha} = 0$. For a generator matrix defined as in Eq. \eqref{eq:generator_matrix} it can be directly shown that $\mu_{\alpha}$ is given by
\begin{equation*}
    \mu_{\alpha}(u)=\frac{1}{Z}d_{\alpha}(u)^2,
\end{equation*}
where $Z$ is a normalization constant \cite{Sarich_2014}. Since the vector $\mu_{\alpha}$ can be calculated directly using the node degrees, IMC offers a reduction of the computational cost compared to LNE, where additional steps for obtaining the transition matrices and matrix comparison are needed.

The IMC approach is capable of detecting changes in the community structure, as it tracks how the invariant measure of community nodes adapts in response to structural changes in the network. As communities are metastable sets of a random walk process, the probability of finding a random walker within a community is in general higher than being outside of it. When the sizes of the communities change over time, e.g., due to a growth or a  shrinkage process, the average invariant measure of community nodes also changes. However, since the invariant measure depends exclusively on node degrees, its ability to capture dynamic network community structure is constrained by the limited information available. For example, if despite the structural change, clusters stay similar in size and density of connections, the IMC method may have difficulties detecting these structural changes. This may be improved in some cases by considering different types of random walk processes \cite{Djurdjevac_2012}. Therefore, it is particularly suitable for datasets with significantly different cluster sizes and node degree distributions in different phases. Since IMC offers a significant reduction in the algorithm's complexity compared to LNE, we will consider it as a special case of LNE and test it on different datasets. When communities change their size, e.g., through merging or splitting, the number of connections within them usually changes, and so does the invariant measure. This is often sufficient to capture the key events that lead to shifts in community structure.

\subsection{Clustering of Time-Snapshots in Temporal Networks: Temporal Random Walk}
\label{sec:temporal_rw}

\begin{figure}[!ht]
  \centering
  \includegraphics[width=1\textwidth]{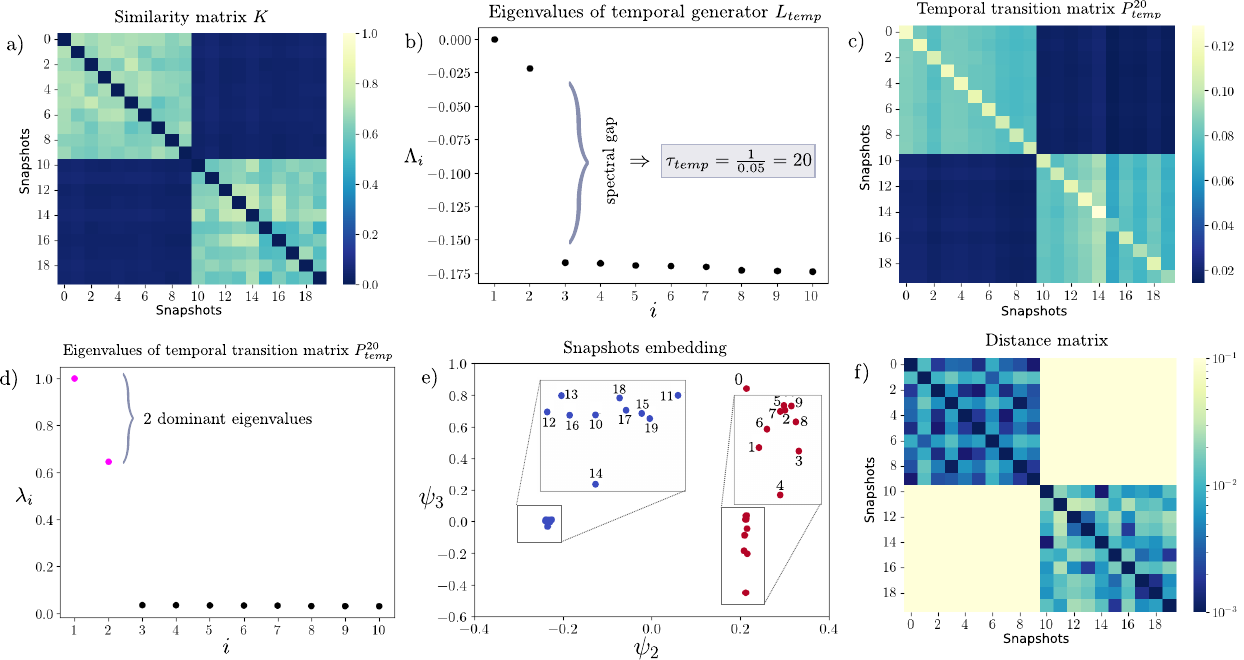}
  \caption{Results of the LNE method applied to the guiding example (see Section \ref{ex:guiding_example}). a) Similarity measures between snapshots, derived from Gaussian kernel evaluations between spatial transition matrices. b) Eigenvalues of the temporal generator $L_{temp}$. c) Temporal transition matrix $P_{temp}^{20}$. d) Dominant eigenvalues of $P_{temp}^{20}$. e) 2-dimensional embedding of entire snapshots partitioned into phases. f) Distance matrix between the rows of matrix $U$.
  }
  \label{fig:LNE}
\end{figure}

In the previous subsection, we introduced transition matrices and invariant measures as objects that can capture crucial information about the internal structure of each time snapshot. Next, our goal is to identify phases during which the global network structure -- particularly the communities within snapshots -- remains largely stable. To achieve this, we employ another time-continuous random walk process at a temporal level, where the random walk transitions are between snapshots and not within them. We define a similarity measure between snapshots by using the Gaussian kernel $k$, which is the standard choice in many applications, although other options are possible \cite{Owhadi_2019}. The Gaussian kernel $k$ is defined by
\begin{equation}
\label{eq:kernel}
    k(P_{\alpha}^{\tau}, P_{\beta}^{\tau})=\exp\left(\frac{-\|P_{\alpha}^{\tau}-P_{\beta}^{\tau}\|_F^2}{2\sigma^2}\right),
\end{equation}
where $\|\cdot\|_F$ denotes the Frobenius norm. The parameter $\sigma$ is called the \textit{bandwidth}. The bandwidth determines the scale of strong relations between snapshots, setting the range within which transition matrices are considered similar. Using these snapshot similarity values, we construct a new static network with similarity matrix $K$ as a weighted adjacency matrix where nodes represent entire snapshots and edge weights $K({\alpha}, {\beta}) := k(P_{\alpha}^{\tau}, P_{\beta}^{\tau})$ reflect the similarity between their community structures (here we set $K(\alpha,\alpha)=0$ for all $\alpha$ to remove loops). This network serves as a reduced model of the initial temporal network. Similarly to Eq. \eqref{eq:generator_matrix} we define a \textit{temporal generator matrix} $L_{temp}$ on this new reduced network. In this matrix, the jump rate from snapshot $\alpha$ to snapshot $\beta$ is given by
\begin{equation}
    L_{temp}(\alpha, \beta)=
    \begin{cases}
        -\frac{1}{d(\alpha)}, & \alpha = \beta, \\[1ex]
        \frac{K({\alpha}, {\beta})}{d(\alpha)^2}, & \alpha \neq \beta,
    \end{cases}
\label{eq:TempGenerator_matrix}
\end{equation}
where $d(\alpha)$ is the (weighted) degree of a node $\alpha$. By selecting an appropriate \emph{temporal exploration parameter} $\tau_{temp}$, with respect to the spectral gap of $L_{temp}$, we obtain the temporal transition matrix $P^{\tau_{temp}}_{temp}$. If multiple spectral gaps are present, different choices of $\tau_{temp}$ allow us to fine-tune the sensitivity of our method when grouping snapshots. Smaller values of $\tau_{temp}$ make the method more sensitive to subtle variations in communities, while larger values filter out milder changes, requiring more substantial differences for phase change detection. In Figure \ref{fig:LNE}(a-d), we show the results of applying these steps on a network from the guiding example (see Section \ref{ex:guiding_example}). Next, we use spectral clustering to group snapshots with consistent community structure into $s$ phases \cite{vonLuxburg_2007}, by computing the matrix $U\in\mathbb{R}^{M\times s}$, whose columns are the eigenvectors corresponding to the dominant eigenvalues of the temporal transition matrix. Finally, we apply $k$-means clustering to the rows of $U$ to detect these phases. The dominant eigenvectors provide low-dimensional embeddings for the snapshots, where snapshots from the same phase are positioned closely in the feature space (Figure \ref{fig:LNE}e). Since the first eigenvector $\psi_1$ is constant and adds no variation to the embeddings, it is typically excluded from the procedure. Therefore, the low-dimensional coordinates of a snapshot~$\alpha$ are given by $(\psi_2(\alpha), \psi_3(\alpha), \ldots)$, where $\psi_2, \psi_3, \ldots$ denote the leading nontrivial eigenvectors. These embeddings are valuable for further temporal network analysis or as input for various machine learning algorithms. Our algorithm is one of the few that can find entire snapshot representations. We will compare its efficacy with alternative methods in Section \ref{sec:related_work_and_comparison}. It is worth noting that the eigenvalue computation for the temporal generator matrix $L_{temp} \in \mathbb{R}^{M\times M}$ is computationally efficient to handle in most practical cases. However, this step could become challenging if an unusually large number of snapshots is considered. For typical applications, $M$ is small enough that this is not a limiting factor.

As an illustration of the results of our algorithm, we show the distance matrix $S\in\mathbb{R}^{M\times M}$ (Figure \ref{fig:LNE}f), whose entries are defined to be distances between rows of matrix $U$:
\begin{equation*}
    S(\alpha,\beta)=\frac{1}{Z}\|(U(\alpha,:)-U(\beta,:)\|,
\end{equation*}
where $Z$ is a suitable scaling parameter. The matrix $S$ is expected to exhibit a near-block structure, with low-value blocks along the main diagonal, indicating closely related snapshots with respect to the chosen kernel. This reflects periods of stability within communities during the evolution of the network.

\begin{figure}[!ht]
  \centering
  \includegraphics[width=1\textwidth]{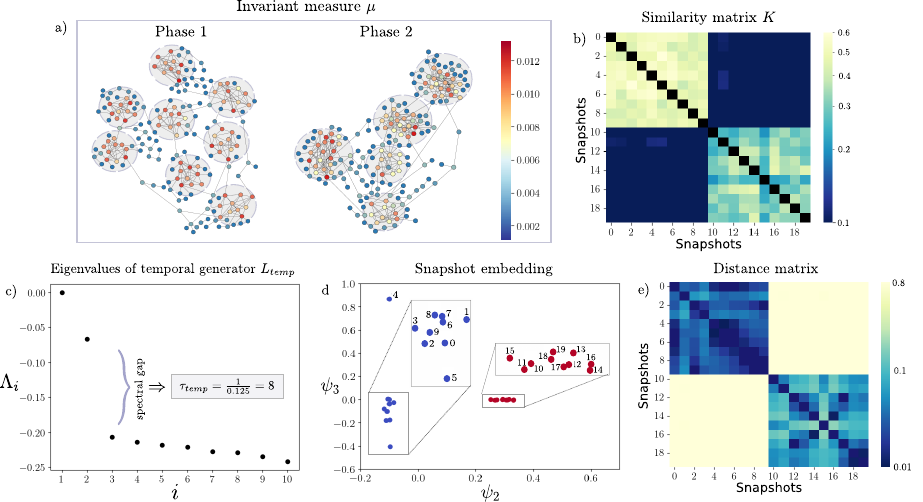}
  \caption{Results of the IMC method applied to the guiding example (see Section \ref{ex:guiding_example}). a) Invariant measure values in base networks. b) Similarity measures between snapshots, derived from Gaussian kernel evaluations between invariant measures. c) Eigenvalues of the temporal generator $L_{temp}$. d) 2-dimensional embedding of entire snapshots partitioned into phases. e) Distance matrix between the rows of matrix $U$.}
  \label{fig:IMC}
\end{figure}

The IMC algorithm is implemented in a similar way, except that we compute $k(\mu_{\alpha}, \mu_{\beta})$ instead of Eq. \eqref{eq:kernel}. By omitting computations of Eq. \eqref{eq:generator_matrix} and Eq. \eqref{eq:P}, this approach achieves greater computational efficiency, although at the cost of some versatility. In Figure \ref{fig:IMC}a we show that communities in the guiding example (see Section \ref{ex:guiding_example}) are easily detectable in the invariant measure values, while in Figure \ref{fig:IMC}b--e we illustrate the IMC method applied on the same example. In this case, both LNE and IMC methods effectively capture the two phases of network evolution, as reflected in the block structure of their respective distance matrices.

\begin{remark}
\label{rem:kernel_adj_comparison}
Note that even though multiple network similarity measures for clustering snapshots have been proposed in the literature \cite{Masuda_2019}, we focus here specifically on comparing snapshots based on their community structure. For example, widely used adjacency matrix-based similarity measures would not be appropriate for this task. Namely, the same set of nodes in two snapshots may exhibit completely different internal wiring patterns such that subnetworks spanning the same node subset could even be complementary, yet still form densely interconnected regions in both snapshots. In this case, the adjacency matrices would differ significantly, even though the overall community structure remains unchanged. Our approach overcomes this issue and effectively identifies consistent communities across snapshots, despite the variations in their internal connections.
\end{remark}

\begin{remark}
\label{rem:diffusion_maps}
We also note that our temporal snapshot embedding bears strong similarity to the diffusion maps dimensionality reduction method  \cite{Coifman_2005}, when applied to snapshots represented via their spatial transition matrices. Specifically, our similarity matrix $K$ serves the role of the affinity matrix in diffusion maps, while the temporal exploration parameter $\tau_{temp}$ is analogous to the diffusion time. In diffusion maps, the $t$-step transition matrix of a Markov chain, denoted $P^t$, can be expressed as
\begin{equation*}
P^t=\sum_{k=1}^M=\lambda_k^t\psi_k\phi_k^T
\end{equation*}
where $\lambda_k$ are the eigenvalues of $P$, sorted in nonincreasing order, and $\psi_k$, $\phi_k$ are the corresponding right and left eigenvectors. Since smaller eigenvalues decay faster over time, one can select an embedding dimension $d$ such that $\lambda_k^t\approx 0$ for all $k>d$. The resulting diffusion map approximately embeds each data point (in our case, snapshot) into a $d$-dimensional Euclidean space via
\begin{equation*}
\varphi_t(G_{\alpha})=[\lambda_1^t\psi_{1,\alpha},...,\lambda_d^t\psi_{d,\alpha}]^T
\end{equation*}
This embedding effectively captures the metastable structure of the random walk, thereby clustering together snapshots that belong to the same phase of the temporal network. In our approach, we choose the parameter $\tau_{temp}$ based on the spectrum of generator matrix $L_{temp}$, ensuring that the $P_{temp}$ exhibits a spectral gap. The associated dominant eigenvectors then yield the desired low-dimensional embedding of the snapshots.
\end{remark}

\subsection{Algorithmic considerations}
\label{sec:identification_of_phases}

Our approach consists of two main steps:
\begin{enumerate}
    \item [(1)] we encode snapshots as transition matrices of time-continuous spatial random walks for a suitably chosen $\tau$;
    \item [(2)] we define a time-continuous temporal random walk across snapshots for a suitably chosen $\tau_{temp}$. Using spectral clustering, we then group these snapshots into phases of community stability.
\end{enumerate}

As discussed above, the choice of $\tau$ is related to the time-scales of the random walk process and its value can be determined from the eigenvalues of the generator. However, the explicit computation of the spectral decomposition of generators for each snapshot is a computationally intensive task. While there is no universal "optimal" value for $\tau$, we provide a 
strategy for its selection  based on network properties. We note that the absence of a principled method for parameter selection is not a limitation unique to our method, but a common challenge in clustering approaches where the number of clusters is not known a priori \cite{Jain_2010, vonLuxburg_2007}. In our case, the choice of $\tau$ can vary depending on several factors, such as:
\begin{itemize}
    \item edge density within communities and average node degree, since longer times are needed to fully explore densely connected communities;
    \item size of communities, where larger communities usually require longer exploration times to fully capture intra-community connectivity; \item irregular structural patterns, may require shorter exploration time to avoid losing fine-grained distinctions.
\end{itemize} In our experiments on real-world and social dynamics simulation datasets, we chose $\tau$ and then verified that this choice aligns with the relevant spectral gap of the generators $L_{\alpha}$. For the temporal random walk, we choose $\tau_{temp}$ based on the spectral gap of the temporal generator $L_{temp}$, since the dimension of these networks is relatively small. This gap reflects how different the phases are, and setting 
the time-scale accordingly ensures that the temporal diffusion is sensitive to transitions between structurally distinct periods while remaining robust to minor fluctuations.

The LNE algorithm is summarized in the following pseudo-code:
\begin{algorithm}[H]
	\caption{Local Neighborhood Exploration (LNE)}
	\begin{algorithmic}[1]
        \State \textbf{Input}: \texttt{adj} -- array of adjacency matrices of a temporal network (alternatively: adjacency lists, incidence matrices), 
        $\tau$ -- spatial exploration time, $\sigma$ -- bandwidth parameter.

		\For {$\alpha=0,1,\ldots,M-1$}
            \State Compute $L_{\alpha}$ and $P_{\alpha}^{\tau}$ using Eq. \eqref{eq:generator_matrix} and Eq. \eqref{eq:P}.
        \EndFor
        \State Initialize empty matrix $K$ of size $M\times M$.
        \For{$\alpha=0,1,...,M-1$}
        \For{$\beta=0,1,...,M-1$}
            \State Set $K(\alpha,\beta):=k(P_{\alpha}^{\tau},P_{\beta}^{\tau})=\exp\left(\frac{-\|P_{\alpha}^{\tau}-P_{\beta}^{\tau}\|^2_F}{2\sigma^2}\right).$
        \EndFor
        \EndFor
        \State Compute $L_{temp}$ using Eq. \eqref{eq:TempGenerator_matrix} for a weighted network with weights given by $K$. Determine $\tau_{temp}$ from the spectral gap of $L_{temp}$. 
        \State Compute $P_{temp}^{\tau_{temp}}$.
        \State \textbf{Apply spectral clustering on} $P_{temp}^{\tau_{temp}}$:
        \State \hspace{2em} Define $U\in\mathbb{R}^{M\times s}$ whose columns are $s$ dominant eigenvectors of $P_{temp}^{\tau_{temp}}$.
        \State \hspace{2em} Run $k$-means on rows of the matrix $U$.
        \State \hspace{2em} Assign the labels to the corresponding snapshots and group them.
	\end{algorithmic}
\label{alg:lne}
\end{algorithm}

Overall, the LNE algorithm combines spectral features from local random walks with global clustering across snapshots, enabling robust phase detection. For sparse networks and moderate number of snapshots, the method remains computationally tractable, with overall complexity dominated by the pairwise comparison step (lines 6-10 in Algorithm \ref{alg:lne}), scaling as $\mathcal{O}(M^2 |V|^2)$, where $M$ is the number of snapshots and $|V|$ is the number of nodes per snapshot. This term arises from computing Frobenius norm distances between $M$ transition matrices of size $|V| \times |V|$, and typically outweighs the costs of transition matrix computation and spectral embedding. (A more detailed analysis has been done by Sarich et al. \cite{Sarich_2014}.)

\begin{remark}
In this work, we compare all possible pairs of snapshots. To reduce the computational complexity, we could also consider a fixed-length sliding window and compute the similarity only between the snapshots within that window. However, this alternative approach would be more sensitive to the occurrence of irregular snapshots (outliers) and would not allow for detecting repeating phases in the network dynamics.
\end{remark}

\section{Experiments on Synthetic Data}
\label{sec:syntdataexp}

Creating and analyzing temporal networks in real-world scenarios present significant challenges, including the time and costs required for data collection and privacy constraints. As a result, relying solely on real-world data for method development and validation is often impractical. Synthetically generated networks provide a valuable alternative, offering a controlled and adaptable testing environment. A well-designed synthetic network generator not only captures the essential dynamics of real-world systems but also allows for the creation of networks with diverse characteristics. This flexibility enables rigorous testing, iterative refinement, and validation of methods in an efficient and reliable manner.

In this section, we first provide an overview of existing benchmark networks and then introduce our novel agent-based temporal network generator for method validation. While existing benchmarks employ various mathematical techniques to emulate real-world dynamics, they are often limited to specific properties or lack flexibility to handle high noise levels and complex behaviors. Our agent-based generator addresses these gaps by offering a versatile approach to simulate diverse scenarios, including community formation, evolution, and metastability in temporal networks. Drawing on well-established agent-based modeling techniques, which are widely used in fields such as social dynamics, disease spreading, and economic systems, our generator creates realistic and adaptable synthetic datasets for robust network analysis.

\subsection{Overview of Existing Benchmark Networks}

Community detection in networks is a multifaceted problem that spans diverse applications, from social networks to biological systems. The diversity and complexity of these systems has driven the development of various benchmarks to evaluate and compare community detection algorithms. These benchmarks aim to replicate key structural and dynamic properties of real-world networks, providing controlled environments for testing. However, given the diversity of network types and behaviors, no single benchmark can address all possible scenarios optimally \cite{Peel_2017}. Established benchmarks often incorporate small-world properties \cite{Watts_1998} and scale-free characteristics \cite{Barabasi_1999}, using network growth and preferential attachment to maintain power-law degree distributions. However, some of the most widely used benchmark networks are based on the planted $l$-partition model \cite{Condon_2001}, such as the Girvan--Newman model \cite{Girvan_2001}, where inter- and intra-community link probabilities are predefined, leading to the formation of distinct clusters. These models, however, often result in networks with homogeneous degree distributions, which do not accurately reflect the complexity of real-world networks. To address this, the LFR benchmark was introduced \cite{Lancichinetti_2008}, incorporating power-law distributions for node degrees and community sizes, as well as a mixing parameter to control the proportion of edges that connect different communities. This approach provides a more realistic representation of the structural diversity observed in real-world networks. Extensions to the LFR benchmark \cite{Lancichinetti_2009} accommodate more complex scenarios, including directed, weighted, and overlapping communities, and have been adapted to dynamic networks (see, e.g., \cite{Lin_2008,Greene_2010}). However, these dynamic adaptations can lead to unpredictable community evolution due to the influence of random processes. To this end, Aldecoa \cite{Aldecoa_2012} introduced a ``closed'' benchmark with predefined initial and final states, where links are rewired gradually to ensure controlled transitions.

Over the last decade, many more temporal network benchmarks have been developed. Notable examples include Granell et al. \cite{Granell_2015}, who proposed cyclic networks based on the stochastic block model (SBM) to simulate events like community splits, merges, growth, and shrinkage, combining these into mixed benchmarks. Rossetti \cite{Rossetti_2017} introduced RDyn, where intra-community relations evolve based on probabilities, with stability measures triggering community events. Bazzi et al. \cite{Bazzi_2019} used probabilistic rules for copying community memberships between snapshots, adding edges via a degree-corrected SBM. Cazabet et al. \cite{Cazabet_2020} employed event instructions (e.g., split, merge, birth, death) to govern community evolution, rewiring edges with the Deterministic Strongly Assortative Block Model (DSABM). Lastly, Longa et al. \cite{Longa_2024} created synthetic surrogate networks mimicking complex system dynamics.

All of these benchmarks aim to replicate the properties observed in real-world dynamical systems as closely as possible. However, accurately modeling all relevant features of such systems becomes increasingly challenging, especially when the systems exhibit significant noise or highly stochastic behavior. This complexity often requires a careful balance between incorporating realistic dynamics and maintaining computational feasibility. Selecting an appropriate benchmark requires careful consideration to ensure it aligns well with the specific characteristics of the system under study, along with proper tuning of parameters to achieve meaningful results.

However, most existing benchmarks are well-suited for evaluating node-level community detection algorithms but are less effective for validating methods that aim to detect global, phase-level structural changes. In particular, many of the aforementioned benchmarks introduce community changes through stochastic rewiring or probabilistic transitions, which limits reproducibility and control over the precise timing and type of transitions. These approaches make it difficult to define clear ground truth phases or to simulate controlled scenarios such as metastable configurations, hierarchical merges, or gradual versus abrupt transitions. To address these gaps, we introduce a new temporal network generator that allows explicit control over the evolution of community structure through a time-dependent potential landscape, enabling the generation of realistic yet fully controllable scenarios with known transition points and customizable noise levels. Our new generator approach aims to complement existing benchmarks by enabling validation at a different level of temporal network organization. The following section describes the details of our new approach.

\subsection{A New Agent-Based Network Generator}
\label{sec:generator}

In this subsection, we introduce a new agent-based temporal network generator designed to address the challenges outlined earlier. Our main idea for obtaining network snapshots is to simulate the dynamics of $N$ agents (also called particles) in a two-dimensional space, 
governed by the diffusion process

\begin{equation*}
    \mathrm{d}X_i(t)=-\nabla U_t(X_i(t)) \, \mathrm{d}t+\sqrt{2\beta ^{-1}} \, \mathrm{d}W_i(t),
\end{equation*}
where $X_i(t)$ is the position of the $i$th agent at time $t$, $\beta$ determines noise intensity, $U_t$ is a time-dependent potential, and $W_i(t)$ is a standard Brownian motion. When simulating stochastic processes using the Langevin equation under the assumption of the fluctuation--dissipation relation, the standard choice for the noise intensity parameter is $\sqrt{2\beta^{-1}}$, where $\beta$ denotes the inverse temperature parameter, ensuring that particles remain in thermal equilibrium \cite{Mattingly_2002, Djurdjevac_2018, Risken_1996}. At each time step, agents represent nodes of a static network, i.e., a snapshot of a temporal network, with edges formed probabilistically based on spatial proximity: the closer two agents are, the higher the probability of an edge between them. In this way, we can synthesize both unweighted and weighted network snapshots. To impose metastability on the benchmark network, we use a time-evolving potential $U_t$ with discrete changes over the simulation time, creating intervals where agents settle near potential minima, forming stable configurations or phases. The shape of $U_t$ leads to the emergence of densely populated regions, resulting in community formation in the network. Changes in the potential perturb the dynamics, leading to the reorganization of agents within the simulation space, which characterizes a shift between two phases. Additional details on the generator’s design and implementation can be found in Section S1 in the Supplementary Information. 

\subsection{Performance Analysis on Synthetic Examples}
\label{sec:performance_analysis}

To validate the performance and robustness of our proposed LNE and IMC methods (introduced in Section \ref{sec:methods}), we conduct experiments on both synthetic and real-world datasets. The following two sub-sections focus on an evaluation based on two synthetic datasets generated using our agent-based temporal network generator. They are designed to highlight different aspects of community dynamics, such as the presence and intensity of noise, density of connections, and the nature of community evolution. In Section \ref{sec:opinions}, we test our algorithm on the underlying network of an agent-based model describing opinion dynamics, in which individuals interact within a social space influenced by media presence and external influencers. In Section \ref{sec:realdataexp}, we will evaluate our method using real-world datasets. A short summary of all used datasets is given in Table \ref{tab:summary}. Additionally, in Section S2 of the Supplementary Information, we test our algorithm on two synthetic null temporal networks without phase transitions to highlight its robustness and reliability (see Figure S1 in the Supplementary Information).

The experiments will be carried out using a range of different exploration times $\tau$ across examples to analyze their influence on detecting structural transitions. Additionally, accurate detection of these transitions relies on the proper choice of the bandwidth parameter $\sigma$ for the Gaussian kernel. We choose $\sigma$ in all experiments using grid search to optimize for the kernel's ability to distinguish between similar snapshots (within the same phase) and dissimilar ones (across different phases). Specifically, we choose $\sigma$ such that the relative variance within the off-diagonal entries of the similarity matrix $K$ is high, enhancing its discriminative power. That is, denoting the off-diagonal elements of $K$ by $\overline{K}$, as a rule of thumb we aim to choose $\sigma$ to be the value that maximizes the ratio $\mathrm{var}(\overline{K})/\mathrm{mean}(\overline{K})$. 

A summary of all parameters  used in the LNE and IMC algorithms across all examples is provided in Table \ref{tab:parameters}. The following subsections present descriptions and analyses of the three synthetic datasets, focusing on their properties and the insights gained from applying LNE and IMC to detect structural transitions and community dynamics.

\subsection{Synthetic dataset 1: Community Split}
\label{sec:Syn_15_120}

\begin{figure}[!ht]
  \centering
 \includegraphics[width=1\textwidth]{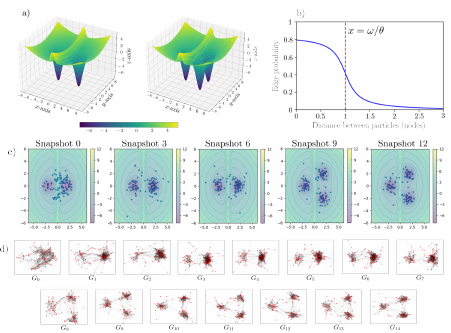}
  \caption{Construction of the \textit{Community Split} dataset using our agent-based network generator. a) Potential function during two equally long periods of the total simulation time. b) Evaluation function for edge construction. c) Positions of the particles and the underlying potential across 5 snapshots.  d) Full realization of a temporal network.}
  \label{fig:generator}
\end{figure}

The \textit{Community Split} dataset is designed to illustrate a straightforward phase transition, where one community splits into two, as shown in Figure \ref{fig:generator}c. We will use this dataset to demonstrate how our methods handle noise and capture changes in community structure over time.

\vspace*{0.2cm}
\noindent \textbf{Dataset Properties}:
The dataset represents the movement of 120 agents over a total simulation time of $T = 7.5$, as shown in Figure \ref{fig:generator}c. The dataset comprises 15 snapshots, created by sampling the simulation every $r = 10$ steps starting from the 10th step. We use the potential function shown in Figure \ref{fig:generator}a that consists of two distinct configurations - or phases. Phase 1 spans snapshots 0-7 and phase 2 corresponds to snapshots 8-14. The parameters defining the potential are selected such that, in the first phase, two wells are present with a barrier separating them, and in the second phase, there are three wells. Snapshots of the agent dynamics can be seen in Figure \ref{fig:generator}c: in phase 1, the simulated agents form two communities localized around the minima of the wells. At the start of phase 2, the right well splits into two, causing the community of agents on the right to also split, inducing the phase transition. The resulting networks presented in Figure \ref{fig:generator}d are constructed based on the agents’ spatial positions, where an edge between two nodes $u$ and $v$ is formed with probability $f(d(u,v))$, where $d$ denotes the Euclidean distance between the nodes. The evaluation function $f$ is defined as
\begin{equation*}
    f(x) = 1 - \xi \left(1 - \nu\left(-\tfrac{1}{\pi} \tan^{-1}(\theta x - \omega) + \tfrac{1}{2} \right)\right),
\end{equation*}
where $\xi, \nu, \theta, \omega \in \mathbb{R}$ are tunable parameters. The parameters $\theta$ and $\omega$ determine the effective distance threshold for community formation, while $\nu$ and $\xi$ control the edge density within and between communities, respectively. The shape of the evaluation function $f$ is illustrated in Figure~\ref{fig:generator}b, and further details regarding this procedure and the specific parameter choices are provided in Section S1 in the Supplementary Information. The snapshots represent intervals of constant potential, ensuring consistent network structure within each phase. The noise intensity is set to $\beta = 0.45$, introducing relatively high variability in agent trajectories while preserving the overall community structure. The snapshots 0 to 7 correspond to Phase 1, and snapshots 8 to 14 represent Phase 2. This transition is initiated by the splitting of the right well, highlighting the generator’s ability to simulate dynamic community evolution (see Figure \ref{fig:generator}d).

\begin{figure}[!ht]
  \centering
  \includegraphics[width=1\textwidth]{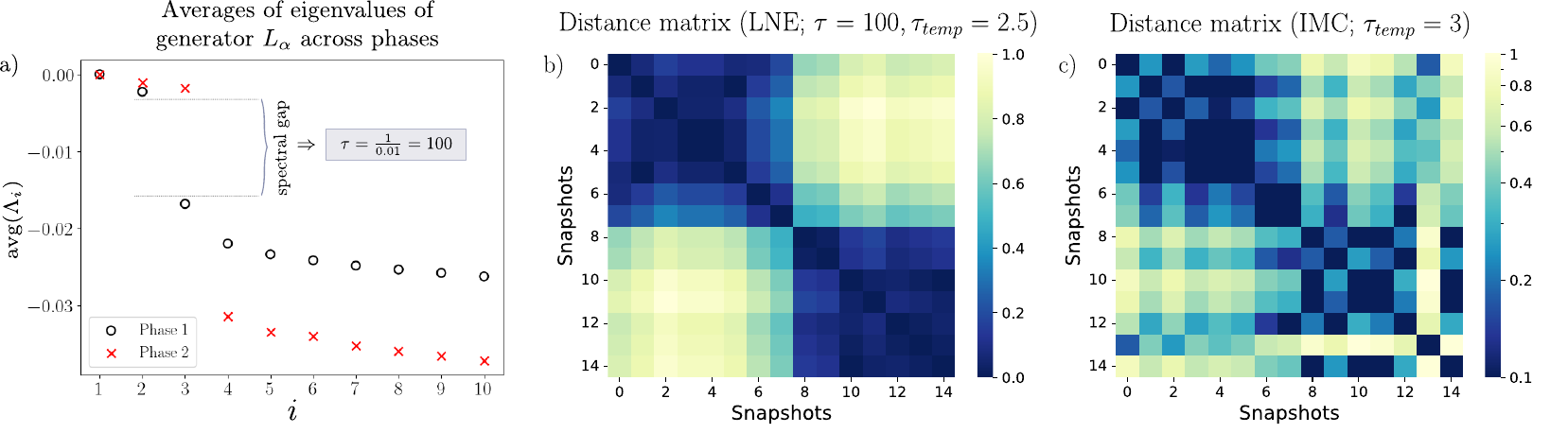}
  \caption{Results of LNE and IMC approaches on the Community Split dataset. a) Averaged eigenvalues of $L_{\alpha}$ across snapshots for both phases. Black circles correspond to snapshots 0-7, red crosses correspond to snapshots 8-14. b) Distance matrix obtained from LNE algorithm for $\tau=100$ and $\tau_{temp}=2.5$. c) Distance matrix obtained from IMC algorithm for $\tau_{temp}=3$.}
  \label{fig:syn_15_120}
\end{figure}

\vspace*{0.2cm}
\noindent \textbf{Results and Observations}:
Spectral analysis of the eigenvalues of the $L_{\alpha}$ generators reveals distinct spectral gaps that correspond to the number of communities in each phase. Specifically, in Phase 1, a gap appears after the second eigenvalue, indicating two communities, while in Phase 2, a gap emerges after the third eigenvalue, reflecting the presence of three communities. We illustrate this behavior in Figure \ref{fig:syn_15_120}a. Since all snapshots within a given phase share a very similar community structure, their eigenvalues remain similar as well, and the spectral gap consistently appears at the same location. Therefore, instead of showing 10 largest eigenvalues for every generator $L_{\alpha}$ for $0\leq\alpha\leq 14$, we show the average of each eigenvalue within Phase 1 ($0\leq\alpha\leq 7$) and Phase 2 ($8\leq\alpha\leq 14$). By doing so, we enhance visual clarity and highlight the general underlying community structure in the snapshots from both phases. Further analysis confirms that for exploration times $\tau$ in the range $1/0.015\approx 67 \leq \tau \leq 200 = 1/0.005$, our method accurately identifies the community structure of both phases. In this example, we set $\tau = 100$ and $\sigma=0.77$. To group snapshots and further characterize phase transitions, we perform spectral clustering on the temporal transition matrix $P_{temp}^{2.5}=\exp({L_{temp}\cdot 2.5})$, which corresponds to a continuous random process between the snapshots. The parameter $\tau_{temp}=2.5$ was chosen based on the spectral gap observed in the spectrum of $L_{temp}$ defined on the reduced network model. The resulting distance matrix (shown in Figure \ref{fig:syn_15_120}b) clearly shows the distinction between the two phases, with a sharp change between the $7$th and $8$th snapshot, corresponding to the splitting of the right community.

In comparison, the IMC method demonstrates lower precision in distinguishing the phases, which is a trade-off for its reduced computational costs. This is also reflected in the less distinct boundaries within the distance matrix, as shown in Figure \ref{fig:syn_15_120}c. Nevertheless, some separation is still visible, with two diagonal blocks corresponding to the two phases. For IMC, we set $\sigma=0.05$ and $\tau_{temp}=3$. Table \ref{tab:comparison} provides a detailed comparison of the performance of LNE and IMC across all examples.

\subsection{Synthetic dataset 2: Community Hierarchy}
\label{sec:Syn_210_100}

The second synthetic dataset, \emph{Community Hierarchy}, is significantly larger than the Community Split dataset, characterized by denser connections within communities and reduced noise. This dataset is designed to demonstrate the hierarchical evolution of community structures across three distinct phases and highlight the limitations of IMC. It exemplifies the generator's ability to simulate complex community dynamics and showcases the method's sensitivity to hierarchical changes, as well as its effectiveness in change-point detection at different resolutions.

\begin{figure}[!ht]
  \centering
  \includegraphics[width=1\textwidth]{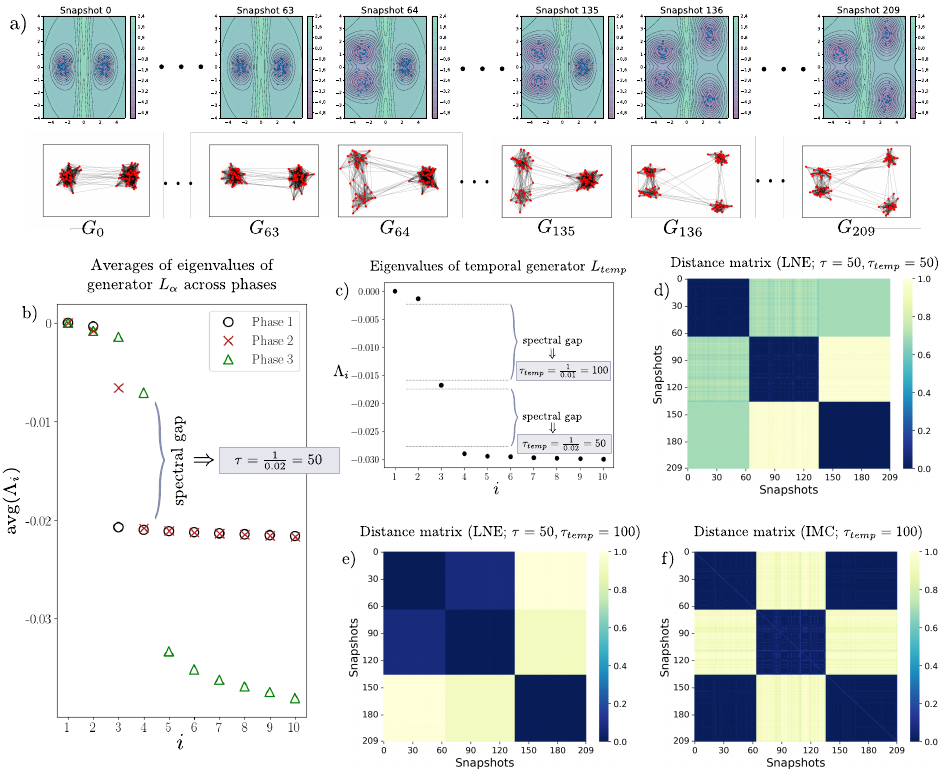}
  \caption{Results of LNE and IMC approaches on the Community Hierarchy dataset. a) Community Hierarchy dataset. b) Averaged eigenvalues of $L_{\alpha}$ across snapshots for all three phases. Black circles correspond to snapshots 0--63, red crosses correspond to snapshots 64--135 and green triangles correspond to snapshots 136--209. c) First ten eigenvalues of the temporal generator $L_{temp}$. d) Distance matrix for LNE, $\tau_{temp}=50$. e) Distance matrix for LNE, $\tau_{temp}=100$. f) IMC, $\tau_{temp}=100$.}
  \label{fig:syn_210_100}
\end{figure}

\vspace*{0.2cm}
\noindent \textbf{Dataset Properties}: The dataset consists of 210 snapshots representing the movement of 100 agents over a total simulation time of $T=105$ with the step size $h=0.05$, where snapshots are sampled every $r=10$ steps. This dataset is divided into three phases of approximately equal duration. In Phase 1, the potential function contains two wells, one on the left and one on the right side of the simulation space, resulting in two communities. At $t=32$ (snapshot $\alpha=64$), the left well splits into two smaller wells, causing the left community to divide into two, marking the start of Phase 2. At $t=68$ (snapshot $\alpha=136$), the right well similarly splits into two, initiating Phase 3. In this case, we set $\beta = 0.75$, reducing the noisy behavior of nodes compared to the Community Split example. The choices of other parameters used to generate this dataset are given in Section S1 in the Supplementary Information and the potential and the resulting network is shown in Figure \ref{fig:syn_210_100}a. Note that the communities on the left maintain closer spatial proximity, resulting in denser connections compared to the communities on the right. This creates a hierarchical structure, where the smaller communities on the left remain part of a larger overarching module.

\vspace*{0.2cm}
\noindent \textbf{Results and Observations}:
Spectral analysis of the eigenvalues of the $L_{\alpha}$ generators shows a clear spectral gap after 2nd, 3rd, and 4th eigenvalue for Phases 1, 2 and 3 respectively. Similarly to the previous example, in Figure \ref{fig:syn_210_100}b averaged eigenvalues across snapshots for all three phases are shown together with the indicated spectral gap. Based on this analysis we choose $\tau=1/0.02=50$ and set $\sigma=0.5$. Further analysis of $L_{temp}$ on the reduced network model reveals two possible spectral gaps in its eigenvalue spectrum (see Figure \ref{fig:syn_210_100}c). Using $\tau_{temp}=100$ based on the first gap, our method detects two phases but does not distinguish the third due to the high interconnectivity of the left-side communities. Reducing $\tau_{temp}$ to $50$, based on the second gap, successfully differentiates all three phases. The distance matrices between snapshots confirm these observations (Figures \ref{fig:syn_210_100}d, e). For $\tau_{temp}=50$ three distinct blocks emerge, corresponding to the three phases. This means that the choice of the temporal exploration time determines the resolution level that is considered during phase clustering and change point detection. With lower $\tau_{temp}$, localized, densely connected regions are treated as distinct modular structures, enabling a fine-grained analysis in the learning process. As the temporal exploration time increases, the walker has sufficient time to traverse somewhat less similar snapshots and the algorithm becomes less sensitive to milder variations within communities that are part of larger modules. This flexibility allows for fine-tuning the algorithm’s sensitivity to the intensity of structural changes, offering an interesting balance between snapshots grouping into phases and change detection precision.

The IMC approach fails to detect the difference in the community structure between the first and the third phase, as illustrated in the distance matrix for $\sigma=0.02$ and $\tau_{temp}=100$ (Figure \ref{fig:syn_210_100}f). Specifically, all communities in both cases are of approximately the same size within snapshots, resulting in very similar invariant measures even though the number, size, and structure of communities are completely different between these two phases.

\subsection{Opinion Dynamics Dataset}
\label{sec:opinions}

In recent years, the field of social dynamics has received substantial attention for its efforts to model the movement of individuals in the social space and investigate their interaction and the subsequent evolution of opinions. Individuals can have various roles in the social space and possess the ability to strongly influence the beliefs of other people and, to a certain extent, also the dynamics of the entire social system. The permanent pillars around which individuals tend to construct their opinions are represented by traditional media, integrated within social platforms. Conversely, influencers, who are private individuals with a significant number of followers, enter and exit the social media space at a significantly faster rate, generate content on trending topics, and modify their behavior and opinions to attract the maximum number of followers. Although they are able to attract audiences at a faster pace than traditional media, their capacity to retain followers in the long term is relatively restricted \cite{Helfmann_2023}.

In this example, we consider an agent-based model of social dynamics based on Helfmann et al. \cite{Helfmann_2023, Helfmann_2023_sup}. This simulation models the opinions of $N=250$ individuals in a 2-dimensional opinion space $D\subset\mathbb{R}^2$ with $L=4$ influencers and $M=2$ traditional media agents. In our simulation, individuals initially adopt opinions uniformly from the set $\{(x,y)|x\in [-7,-4]\cup [4,7], y\in [-2,-0.5]\cup [0.5,2]\}$ consisting of four disjoint  regions. Each individual within a region subscribes to the designated influencer and each influencer adopts the average opinion of their respective followers. Positioned initially at $(0,-1.5)$ and $(0,1.5)$ media agents are followed by individuals positioned on the corresponding side of the $x$-axis. During the simulation, individuals interact with each other based on the underlying relational network (in this example, the relational network is complete) and each of them follows exactly one influencer and one medium. The opinion dynamics of individuals is governed by a stochastic differential equation, which includes interaction forces from other individuals, media, and influencers. Influencers adjust their opinions based on the average opinion of their followers, while media agents update their opinions at a slower rate (see Helfmann et al. \cite{Helfmann_2023} for details and default parameter values). The simulation is conducted over 5000 iterations where snapshots are taken with a step size of 30 iterations. A weighted temporal network is then constructed over the set of individuals, where the edge weights at each snapshot are given with $w_{ij}(t)=\exp(-|x_j(t)-x_i(t)|)$ where $x_i(t)$ represents the opinion of individual $i$ at time $t$ (following the original approach).

\begin{figure}[!ht]
  \centering
  \includegraphics[width=1\textwidth]{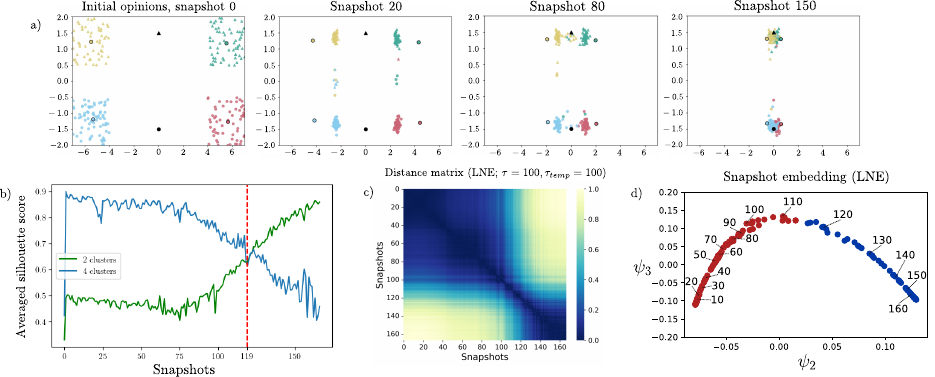}
  \caption{Results of the LNE approach applied to the Opinion Dynamics dataset. a) Initial positions of agents. Individuals grouped around 4 influencers gravitate towards 2 media outlets, ultimately forming only 2 distinct communities. b) Averaged silhouette scores over 5 iterations for 2- and 4- clusterings. c) LNE distance matrix for $\tau=100$ and $\tau_{temp}=100$. d) Clustered embeddings of the Opinion Dynamics dataset.}
  \label{fig:opinions}
\end{figure}

After applying LNE to this temporal network, we recover two phases of the system as illustrated by the distance matrix in Figure~\ref{fig:opinions}c and the snapshot embedding space in Figure \ref{fig:opinions}d. The first phase corresponds to the period preceding the merging of communities, while the second one corresponds to the period when all individuals belong to one of the two distinct communities. Since the ground truth phases of this dataset are missing, we identify them by comparing the silhouette scores of 4- and 2-clusterings \cite{Rousseeuw_1987}. Intuitively, the silhouette score measures how well a node fits within its assigned community (cohesion) relative to other communities (separation). We define the transition point between the phases with 4 and 2 communities as the first snapshot where the average silhouette score for 2-clustering, computed over 5 iterations, exceeds that for 4-clustering (Figure \ref{fig:opinions}b). All clusterings of individuals are performed using $k$-means. A visual representation of four snapshots showing the current positions of agents across various time points is given in Figure \ref{fig:opinions}a. The black circle and triangle represent two media sources, influencers are illustrated as colored circles, and individuals take on the shape of the media they follow and adopt the color of the influencer they are subscribed to. Initially, individuals are randomly distributed within designated regions of the opinion space. Influencers adopt the average opinion of their followers, so the individuals form four communities in the vicinity of the corresponding influencer. They gradually gravitate towards the media outlet they follow until the communities in the upper and lower half-planes of the social space start to merge. Ultimately, individuals are divided into two discernible communities. In this dataset, we set $\tau=100,$ $\sigma=0.66$ and $\tau_{temp}=100$ in LNE. As mentioned in Section \ref{sec:IMC}, the IMC algorithm encounters difficulties in distinguishing between phases due to the similar sizes of communities within all snapshots. For this experiment, we set $\sigma = 0.01$ and $\tau_{temp} = 10$. This example is of particular interest because all the snapshots are complete networks with different interaction strengths (weights) between the nodes. As shown in Table \ref{tab:comparison} most of the methods had difficulties in handling this gradual transition period and detecting a clear change point. Our approach and GraphKKE (Graph Kernel Koopman Embedding) achieved notably better performance. Note that GraphKKE works only with unweighted networks (see Remark \ref{rem:GraphKKE}) so in this case we adapt the dataset by introducing an edge cut-off threshold to get the best outcome (see Section \ref{sec:comparison} for details).

\section{Experiments on Real-world Datasets and Social Dynamics Simulation Data}
\label{sec:realdataexp}
Having demonstrated the effectiveness of our method on synthetic datasets, we now turn our attention to more complex and real-world scenarios. In this section, we apply our approach to real data, showcasing its practical applicability across various domains of network analysis. Real-world datasets often present unique challenges, such as incomplete data, noise, and complex, domain-specific dynamics. To demonstrate the practical utility of our approach, we apply it to diverse datasets spanning biological and sociological contexts. Specifically, we will use the following three datasets, which are also listed in Table \ref{tab:summary}: (1) Vibrio Cholerae Infection  \cite{Hsiao_2014, Melnyk_2020} (real-world dataset of gut microbiota parameters of a Bangladeshi patient during the disease and recovery period), (2) a protein--protein-interaction network during cell division \cite{Lucas_2023} (temporal network modeling protein--protein interactions during the yeast cell division cycle) and (3) Primary School Contacts network (temporal network modeling face-to-face contacts between primary school students during one working day.) We should note that in the Cholera and Primary School Contacts datasets we have several connected components or isolated nodes within snapshots. In such cases we treat each connected component independently and consider the probability distributions resulting from random walks on each component individually. The generator matrix $L$ in this case has zero rows corresponding to isolated nodes. In these examples, we select $\tau$ as mentioned in Section \ref{sec:identification_of_phases}. Table \ref{tab:summary} presents the averages of the expected waiting times at nodes across snapshots. More precisely, the 
choice of $\tau$ is based on the ratio $\tau$/(avg. degree), which reflects the expected number of steps of a random walker during the exploration phase. For datasets with dense community structure, such as the Opinion Dynamics dataset from Section \ref{sec:opinions}, a small ratio (1.44 in this case) is sufficient to capture the local structure effectively. In contrast, for datasets with sparser intra-community connections, such as the Cholera dataset, a higher ratio (13.5) is selected to ensure adequate exploration of the community structure.  The parameter $\sigma$ is determined as described in Section \ref{sec:performance_analysis}, while $\tau_{temp}$ is directly computed from the spectral gap of $L_{temp}$. A list of the used parameters for all experiments is shown in Table \ref{tab:parameters}.

{\scriptsize
\begin{longtable}{@{}l c c c c c}
\caption[]{Summary of datasets used for the benchmark experiments} \\
\toprule%
\centering%
 & \multicolumn{1}{c}{{{\bfseries \#Nodes}}}
 & \multicolumn{1}{c}{{{\bfseries Avg. \#Edges/Edge weight}}}
 & \multicolumn{1}{c}{{{\bfseries Avg. degree}}}
 & \multicolumn{1}{c}{{{\bfseries \#Snapshots}}}
 & \multicolumn{1}{c}{{{\bfseries \#Phases}}} \\

\cmidrule[0.4pt](r{0.125em}){1-6}%
\endfirsthead

\myrowcolour%
Community Split & 120 & 1137.2 & 18.95 & 15 & 2 \\

Community Hierarchy & 100 & 1551.3 & 31.03 & 210 & 2/3 \\

\myrowcolour%
Opinion Dynamics & 250 & 8666.83 & 69.34 & 167 & 2 \\

Cholera & 96 & 35.58 & 0.74 & 34 & 2 \\

\myrowcolour%
Cell Division & 83 & 135.65 & 3.26 & 102 & 2-11\\

Primary School Contacts & 238 & 934.89 & 10.41 & 103 & 3 \\

\bottomrule
\label{tab:summary}
\end{longtable}
}

\subsection{Cholera Dataset}
\label{subsec:cholera}

\begin{figure}[!ht]
  \centering
  \includegraphics[width=1\textwidth]{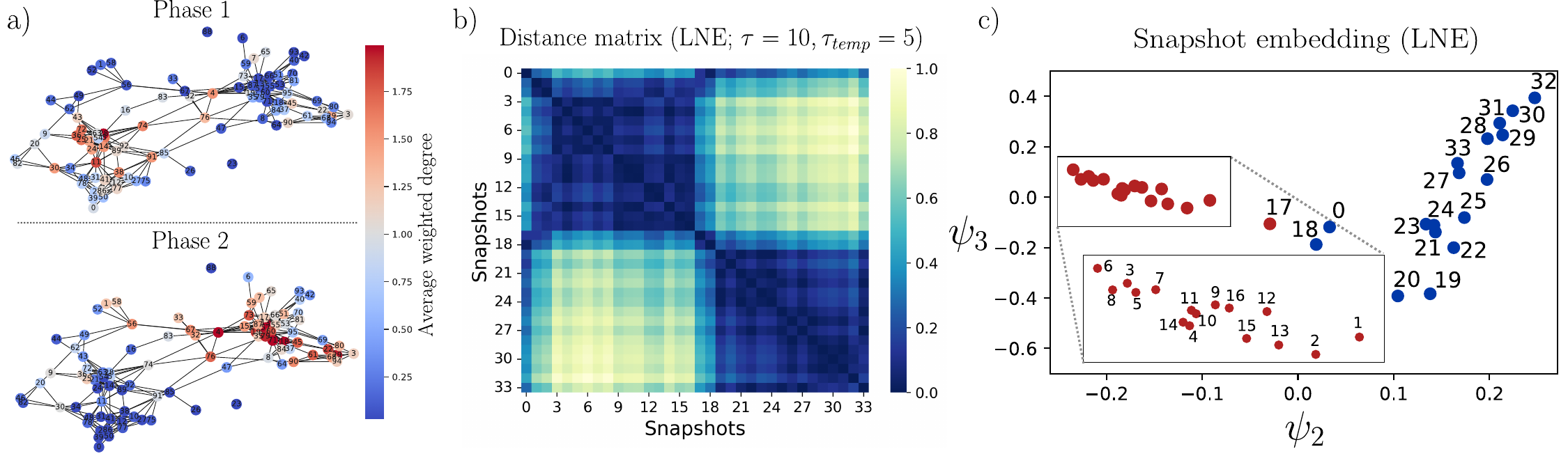}
  \caption{Results of the LNE approach applied to the Cholera dataset. a) Average weighted degree of nodes in the OTU interaction network across two distinct phases. b) LNE distance matrix for $\tau=10$ and $\tau_{temp}=5$. c) Clustered snapshot embedding using the second and third eigenvector of the temporal transition matrix.}
  \label{fig:cholera}
\end{figure}

The first dataset that we will analyze originates from research conducted in Hsiao et al. \cite{Hsiao_2014} (see Melnyk et al. \cite{Melnyk_2020} for an analysis of this data). The focus of this study was to understand the changes in the gut microbiota during the course of Vibrio cholerae infection. The data was collected from a Bangladeshi patient over time, during different stages of the disease and convalescence period. Based on the total of 34 collected samples, a temporal network is constructed as described in Melnyk et al. \cite{Melnyk_2020} modeling the presence and correlations between operational taxonomic units (OTUs) during both disease manifestation and recovery phase. The average weighted degree of nodes in the OTU interaction network across the two phases is illustrated in Figure \ref{fig:cholera}a. The results indicate the formation of different communities in each phase, corresponding to the patient's health status. Our objective is to identify these two phases within the OTU data, as well as to detect the moment in time when the transition between infection and recovery period happened. Figure \ref{fig:cholera}b shows the distance matrix obtained from LNE with $\tau=10$, $\sigma=2.75$, and $\tau_{temp}=5$. The original study reports that the patient entered the recovery phase after snapshot 17, which closely aligns with the results of our method. The embeddings obtained from the second and third eigenvector of the temporal transition matrix are shown in Figure \ref{fig:cholera}c, where data points are colored based on spectral clustering results. For the IMC experiments given in Table \ref{tab:comparison} we set $\sigma=0.03$ and $\tau_{temp}=2$.

\subsection{Cell Division Dataset}
\label{sec:cell_division}

\begin{figure}[h!]
  \centering
  \includegraphics[width=\textwidth]{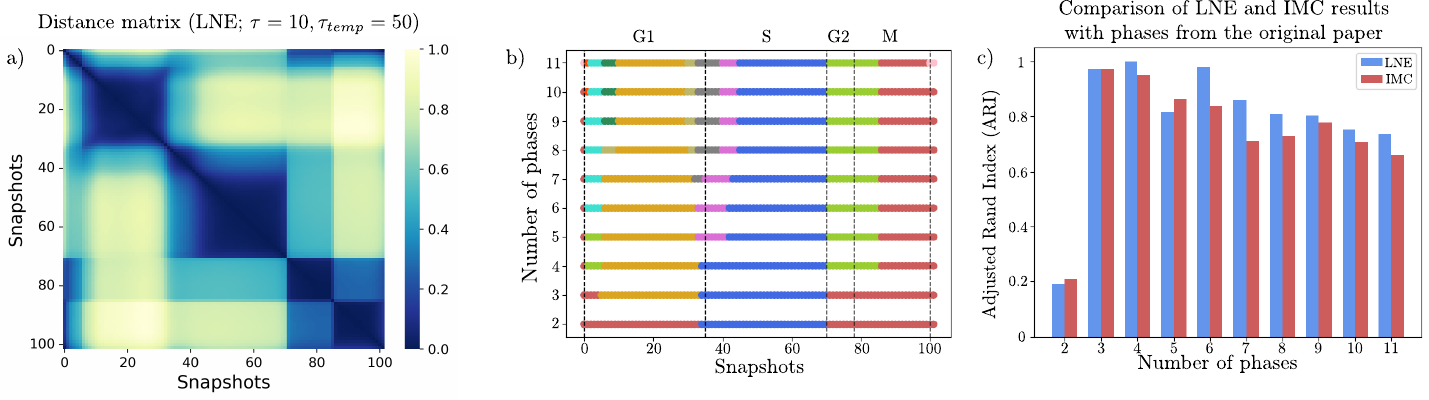}
  \caption{Results of LNE and IMC performance on the Cell Division dataset. a) LNE distance matrix for $\tau=10$ and $\tau_{temp}=50$. b) Phases inferred for $k$ ranging from 2 to 11. c) Comparison of LNE and IMC results with the clusterings from the original paper.}
  \label{fig:cells}
\end{figure}

In this example, we consider a biological real-world dataset from a study by Lucas et al.  \cite{Lucas_2023} on yeast cell division cycles. Simply put, cells undergo four main biological phases before dividing into two: first, cells grow to the necessary size in the first phase (G1), followed by DNA synthesis (S phase), a second gap phase (G2), and finally mitosis (M phase), where duplicated chromosomes are evenly divided into two daughter cells. Cell division is regulated through the interactions between proteins, which can be modeled as a protein-protein interaction (PPI) network \cite{Lucas_2023, Newman_2010}. In PPI networks, nodes represent proteins, whereas the weighted edges represent the interactions among them. To model the temporal network of protein interactions during the yeast cell cycle, data from the static PPI network are enhanced with temporal series data on edge activity. This enhancement involves numerical derivations of protein concentrations over time, based on an existing ODE model that describes the interactions for approximately 21\% of the proteins  \cite{Chen_2000}. The granularity of the time snapshots of the network is dictated by the time step intervals used in the ODE's numerical integration. Evolving edge weights are calculated using available data on known protein-protein interactions, representing the simultaneous presence of corresponding proteins at specific times (and are subsequently normalized). Lastly, weights for the remaining edges from the static PPI network are uniformly assigned a value of 1. Due to the changes in protein-protein interactions following transitions between different phases of a cell cycle, the structure of network snapshots should exhibit consistency within the same phase, while significant changes in the network structure are expected to occur following a phase transition. Lucas et al. \cite{Lucas_2023} developed the \emph{Phasik} algorithm that allows to detect $k$ groups of structurally similar snapshots for a predefined parameter $k$. They obtained 10 partitions of the network for $k$ ranging from 2 to 11. Although this dataset is not expected to exhibit pronounced community structures, we apply our LNE and IMC algorithms here as well and compare the results with clusterings from the original paper \cite{Lucas_2023} to test their performance in a more generalized setting (Figure \ref{fig:cells}). In this experiment we set $\tau=10, \sigma=0.6$, $\tau_{temp}=50$ in LNE and $\sigma=0.01$ and $\tau_{temp}=50$ in IMC. The results demonstrate that the phases we detected align closely with those reported by the original authors, particularly for smaller values of $k$. One exception is the case of $k=2$, where our method isolated only the S phase from the others, whereas the original algorithm grouped the G1 and S phases together, separating them from the G2 and M phases. Additionally, we observe that the transition between the G2 and M phases occurs later in the network's evolution across all $k$. This phenomenon, also noted in the original results, is attributed to a sharp change in edge weights between specific protein interactions. Biologically, this corresponds to the metaphase checkpoint, where chromosomes are expected to achieve proper bipolar attachment to the mitotic spindle.

\subsection{Primary School Contacts Dataset}
\label{sec:primary_school}

\begin{figure}[h!]
  \centering
  \includegraphics[width=0.85\textwidth]{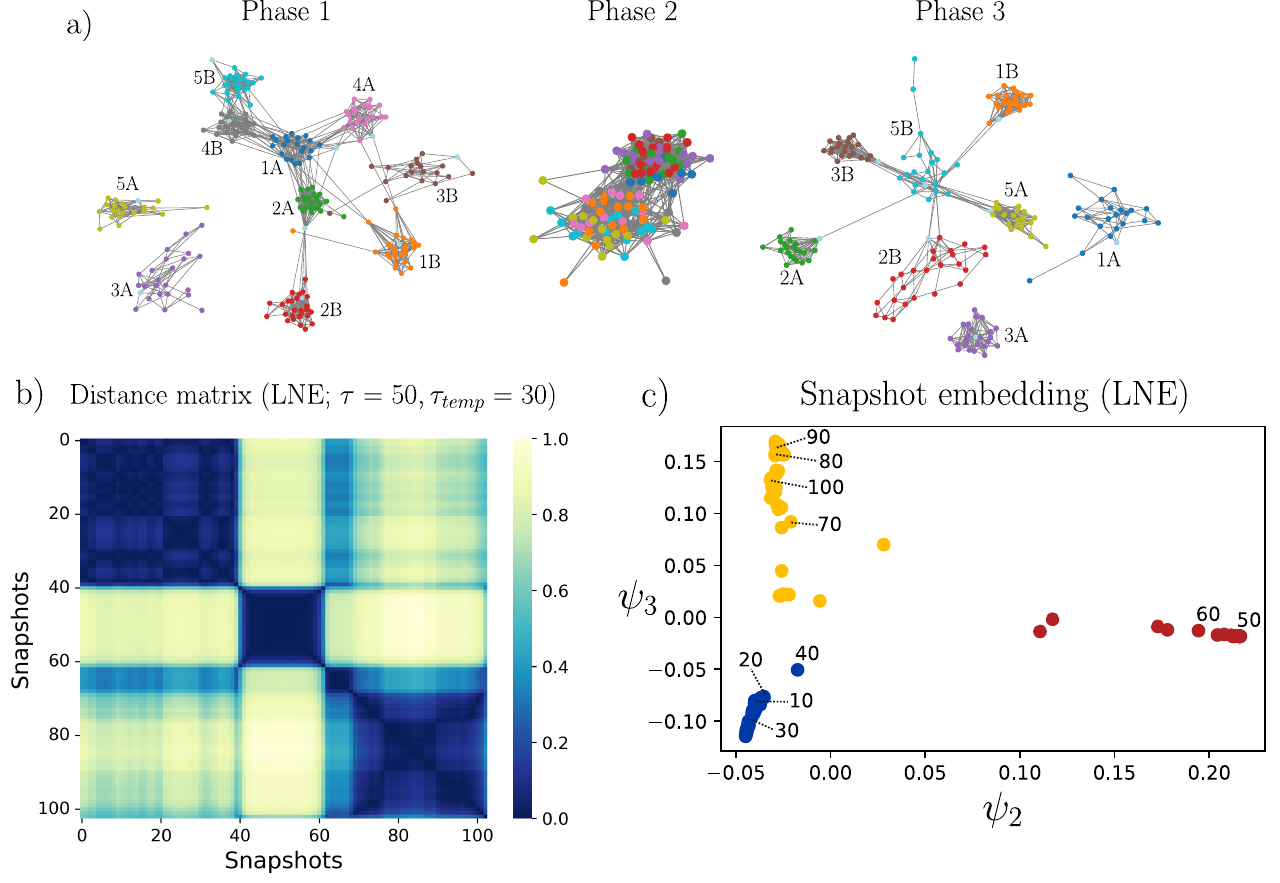}
  \caption{Snapshots of the temporal network across three periods of the day: morning shift (Phase 1), lunch break (Phase 2), and afternoon shift (Phase 3). Light blue nodes within each class represent teachers. b) LNE distance matrix for $\tau=50$ and $\tau_{temp}=25$. c) Clustered embeddings of the Primary School Contacts dataset.}
  \label{fig:school}
\end{figure}

As a last example, we analyze a face-to-face contact dataset collected from primary school students over two working days, which is publicly available at \url{www.sociopatterns.org}. The dataset records contacts between students from grades 1 to 5, with each grade divided into two classes, labeled A and B. Data collection was conducted by Stehlé et al. \cite{Stehle_2011} in a primary school in Lyon, France, on the 1st and 2nd of October 2009. Each participant was equipped with a Radio-Frequency Identification (RFID) badge, worn throughout the day. The badges were configured to detect face-to-face interactions, recording a contact when two individuals were within 1 to 1.5 meters of each other. RFID devices exchanged one radio packet every 20 seconds, and a contact was considered established only if individuals remained in proximity long enough for at least one packet exchange to occur. Contacts were considered ongoing as long as successive packet exchanges continued without a gap exceeding 20 seconds. At each 20-second interval, a network snapshot was constructed: nodes represent students, and edges indicate contacts between them. In total, the data includes interactions among 232 students and 10 teachers. The school day is divided into three parts: the morning shift, the lunch break (from 12 PM to 2 PM), and the afternoon shift. Stehlé et al. \cite{Stehle_2011} concluded that during classes (the morning and afternoon shifts), students, as expected, primarily interact with their classmates and teachers, which is clearly noticeable in Figure \ref{fig:school}a, where nodes are colored according to class membership and communities accurately correspond to classes. However, during the lunch break, many students go home, and those who remain mix together in the cafeteria, leading to interactions that spread beyond class boundaries. As a result, the clear division of the network into communities is lost, and the remaining students form a single large community. In our experiment, we consider a temporal network constructed from data corresponding to a single working day. We select October 2 due to the overall higher intensity of contacts observed on that day, as noted in  \cite{Stehle_2011}. Individual snapshots, recorded over 20-second intervals, are naturally quite sparse; therefore, we form our dataset by aggregating contacts between students over 30-minute intervals, with each interval starting 5 minutes after the previous one. The resulting temporal network consists of 103 snapshots. Additionally, we note that after the lunch break, classes 4A and 4B left the school, leading to the disappearance of the corresponding communities in subsequent snapshots. In Figure \ref{fig:school}a, we show one snapshot from each part of the day, displaying only nodes that have at least one connection. 

We applied our method to this example and set $\tau=50, \sigma=4, \tau_{temp}=30$ in LNE and $\sigma=0.05, \tau_{temp}=25$ in IMC. 
We observe that both LNE and IMC methods successfully detect three phases, corresponding precisely to the three periods of the day, see the LNE distance matrix in Figure \ref{fig:school}b and Table \ref{tab:comparison}. The morning and afternoon shifts are identified as distinct phases, due to the absence of two communities -- classes 4A and 4B -- in the latter. Since we use a 30-minute sliding window to aggregate snapshots, we define the ground truth assignment by stating that a snapshot belongs to a given phase (morning shift, lunch break, or afternoon shift) if at least half of the time captured by the snapshot falls within that period. Specifically, for boundary cases between the morning/afternoon shifts and the lunch break (where 15 minutes fall into each phase), we classify these snapshots as part of the lunch phase, as connections are overall significantly denser during that period. In Figure \ref{fig:school}c, we show the snapshot embeddings, where the three previously mentioned phases are clearly distinguishable.

\begin{longtable}{@{}l c c c | c c@{}}
\caption[]{Parameter values for our LNE and IMC methods, used in the benchmark experiments.} \\
\toprule
\centering%
 & \multicolumn{3}{c}{{\bfseries LNE}} & \multicolumn{2}{c}{{\bfseries IMC}} \\
\cmidrule[0.4pt](r{0.125em}){2-4} \cmidrule[0.4pt](r{0.125em}){5-6}
 & $\tau$ & $\sigma$ & $\tau_{\text{temp}}$ & $\sigma$ & $\tau_{\text{temp}}$ \\
\midrule
\endfirsthead

\toprule
\centering%
 & \multicolumn{3}{c}{{\bfseries LNE}} & \multicolumn{2}{c}{{\bfseries IMC}} \\
\cmidrule[0.4pt](r{0.125em}){2-4} \cmidrule[0.4pt](r{0.125em}){5-6}
 & $\tau$ & $\sigma$ & $\tau_{\text{temp}}$ & $\sigma$ & $\tau_{\text{temp}}$ \\
\midrule
\endhead

Community Split & 100 & 0.77 & 2.5 & 0.05 & 3 \\
\rowcolor[HTML]{EFEFEF}
Community Hierarchy (2 phases) & 50 & 0.5 & 100 & 0.02 & 100 \\
Community Hierarchy (3 phases) & 50 & 0.5 & 50 & 0.02 & 100 \\
\rowcolor[HTML]{EFEFEF}
Opinion Dynamics & 100 & 0.66 & 100 & 0.01 & 10 \\
Cholera & 10 & 2.75 & 5 & 0.03 & 2 \\
\rowcolor[HTML]{EFEFEF}
Cell Division & 10 & 0.6 & 50 & 0.01 & 50 \\
Primary School Contacts & 50 & 4 & 30 & 0.05 & 25 \\
\bottomrule
\label{tab:parameters}
\end{longtable}

\section{Benchmarking Community Detection in Temporal Networks: A Comparative Review}
\label{sec:related_work_and_comparison}

As discussed above, analyzing community dynamics in temporal networks presents unique challenges. Methods are required that can accurately detect structural transitions while maintaining computational efficiency. The growing interest in this area has led to the development of diverse approaches, ranging from spectral clustering to machine learning-based embeddings. To provide the reader with a clearer understanding of the field, we first summarize some of the existing key methods and then present a comparative analysis of these methods with our methods.

\subsection{Methods for Community Detection in Temporal Networks}
\label{sec:methods_for_community_detection}

Numerous algorithms have been developed for community detection in static and temporal networks. While static methods like modularity optimization \cite{Newman_2004} and spectral clustering \cite{vonLuxburg_2007} are well-established, their extensions to temporal networks face challenges due to the dynamic nature of evolving communities \cite{Huang_2021}. Approaches tailored to temporal networks often aim to identify structural similarities across time points and to detect significant changes, such as community splits/merges, births/deaths, or complete reorganizations. 
Multiple approaches have been proposed to detect change points in temporal networks. For example, Barnett and Onnela \cite{Barnett_2016} focus on identifying the time point at which the difference between covariance matrices of node features before and after the candidate change point is maximized. On the other hand, Peel et al. \cite{Peel_2014} employ probabilistic models -- such as generalized hierarchical random graphs -- combined with Bayesian hypothesis testing to determine when a change point occurs. Additionally, transfer operator methods \cite{Melnyk_2020,Klus_2022} leverage spectral properties of stochastic processes to capture transitions in network structure. Several community detection methods originally designed for static networks have also been extended to temporal settings. Palla et al. \cite{Palla_2007} built on earlier work \cite{Palla_2005} using the autocorrelation function to quantify the overlap of communities across consecutive snapshots. Random walk-based methods, such as those exploiting metastability of Markov chains within communities \cite{Delvenne_2010}, have also been generalized to account for temporal dynamics. Bovet et al. \cite{Bovet_2022} proposed a time-dependent random walk constrained by the presence of edges at different times during the evolution and derived a quality function that identifies partitions corresponding to the most persistent communities over time. Peixoto and Rosvall \cite{Peixoto_2017} developed a data-driven method that fits networks to a Markov chain model, enabling the inference of community structures from the real-world data. Recent approaches by Li et al.   \cite{Li_2023,Li_2024} incorporate node attributes via a cluster potential game and dynamic belief systems to detect overlapping communities in attributed networks. Lastly, Rosvall and Bergstrom \cite{Rosvall_2010} extended the their well-established map equation framework \cite{Rosvall_2008}. By incorporating bootstrap resampling and significance clustering, their approach allows for the detection of community changes over time.

A key direction in temporal network analysis is the development of methods that learn low-dimensional embeddings of network snapshots, capturing their structural information and enabling the detection of (meta-)stable states and transitions in this new feature space. While much of the existing work focuses on embedding nodes or edges in static networks \cite{Perozzi_2014, Grover_2016, Gao_2019}, fewer approaches address the challenge of embedding entire snapshots for temporal networks \cite{Holme_2012, Mahdavi_2018}, or even temporal networks as a whole \cite{DallAmico_2024}. This remains an active area of research, with significant potential for applications in analyzing evolving systems. Next, we outline several key methods that aim to derive low-dimensional representations of snapshots and discuss their comparative performance in Section \ref{sec:comparison}.

\vspace{0.2cm}

\textit{Principal Component Analysis (PCA)}. PCA is a standard dimensionality reduction algorithm that discovers new uncorrelated variables, also called principal components, in large, high-dimensional datasets such that only a first few of them contain as much information as possible about the system. In other words, it creates a new coordinate system such that most variance in the data is aligned with the direction of the first principal component. Each subsequent component captures the maximum remaining variance while being orthogonal to the previous components. We will apply PCA on the set of time-snapshots adjacency matrices.

\textit{Graph Kernel Koopman Embedding (GraphKKE)} employs transfer operators to detect structural changes in temporal networks  \cite{Melnyk_2020}. These operators, such as adjoint Perron–Frobenius and Koopman operators, describe the evolution of probability densities and observables in stochastic processes. While these operators are infinite-dimensional, their spectral properties can be approximated using, e.g., kernel methods. For temporal networks, snapshot similarities are captured using kernel evaluations between snapshots and their time-lagged versions. The resulting spectral gap in the eigendecomposition reveals the number of metastable states, with eigenfunctions providing low-dimensional embeddings of the snapshots.

\textit{Transformer + Contrastive Learning. (T+CL)} is an unsupervised transformer-based approach \cite{Melnyk_2023}. Each snapshot is represented by a ``master node'', connected to all other nodes and assigned a learnable vector. The method employs extended adjacency matrices as masks in multi-head attention layers to capture the topological structure of the snapshots. The algorithm computes node embeddings, where master node embeddings serve as low-dimensional representations of the entire snapshots. Additionally, the method considers the temporal order of snapshots by using the embeddings of several preceding snapshots as part of the input information while learning the next representation. Contrastive learning is then applied to group structurally similar snapshots closely in the embedding space while separating those from different phases, using the InfoNCE \cite{Oord_2018} loss function.

\textit{tdGraphEmbed} integrates temporal information to create low-dimensional representations of entire snapshots \cite{Beladev_2020}. This approach uses an unsupervised natural language processing (NLP) technique with a \emph{word2vec} based architecture \cite{Mikolov_2013}. The algorithm utilizes information collected via random walks implemented as in Grover et al. \cite{Grover_2016} that incorporate tunable hyperparameters $p$ and $q$ to regulate the \emph{Breadth-First Search} (BFS) and \emph{Depth-First Search} (DFS) biases of the walks. These walks collectively form a document that encodes the network's snapshot at that specific time. Additionally, a learnable vector representing the entire snapshot $G_{\alpha}$ is given, and the goal is to maximize the likelihood of observing $G_{\alpha}$ given the corresponding document.  This extends the Continuous Bag of Words (CBOW) model to encode both node relationships and snapshot-level features.

Together, these methods demonstrate a range of techniques for analyzing temporal networks by embedding snapshots in low-dimensional spaces. Their comparative evaluation, discussed in the next section, provides insights into how these approaches perform across synthetic and real-world datasets.

\subsection{Comparative Analysis}
\label{sec:comparison}

In this section, we will compare the performance of our LNE and IMC approaches to the algorithms described in the previous section, showing their performance on the introduced synthetic and real-world datasets. We evaluate the performance of our algorithm and others on these datasets using the adjusted Rand index (ARI) and normalized mutual information (NMI). ARI \cite{Hubert_1985} measures the agreement between predicted and ground truth clusterings by counting how consistently pairs of elements are assigned to the same or different clusters, while correcting for agreement due to chance. On the other hand, NMI \cite{Fred_2003} quantifies how much information is shared between two clusterings, capturing how well the structure of one clustering predicts the structure of the other. We set the following method-specific parameters. In all examples, we set the number of components in PCA to 2. In GraphKKE, we set the number of iterations $h=3$ for Primary School Contacts dataset and $h=1$ for all others, regularization parameter $\eta=0.1$ and lag time $\tau=1$. Since GraphKKE accepts only unweighted networks (see Remark~\ref{rem:GraphKKE}), where edges can only have weights 0 or 1, we used grid search to identify the best edge cut-off parameter. Accordingly, we preprocessed the Opinion Dynamics dataset by removing all edges with weights below 0.15, and assigning a weight of 1 to the remaining edges. In the Cholera dataset, as in the original paper, all edges with positive weight are kept and assigned a new weight equal to 1. In T+CL we set batch sizes to 15 for Community Split, 64 for Community Hierarchy, Opinion Dynamics and Primary School Contacts and 6 for Cholera. The rest of the parameters are set at default values as in Melnyk et al. \cite{Melnyk_2023} In tdGraphEmbed we set $p=1$ and $q=6$ so that random walks used for representation learning to explore their originating communities. The number of walks simulated from each node of each snapshot is set to 40 and their length to 50. The learning process is carried over 50 iterations, and the dimension of the embedding space is 128. To allow the tdGraphEmbed algorithm time to learn, several initial embeddings are excluded from the ARI and NMI computations.

The following publications and codes were used to check and test the performance of the methods described in Section \ref{sec:related_work_and_comparison} on our datasets:
\begin{enumerate}
\item \textit{Principal Component Analysis (PCA)}: PCA module of the \texttt{scikit-learn} library.
\item \textit{Graph Kernel Koopman Embedding (GraphKKE)}: \cite{Melnyk_2020} and the code from GitHub repository \cite{graphKKE_git_2020}.
\item \textit{Transformer + Contrastive Learning (T+CL)}: \cite{Melnyk_2023} and the code from GitHub repository \cite{deep-metastability_git_2023}.
\item \textit{tdGraphEmbed}: \cite{Beladev_2020} and the code from GitHub repository \cite{tdGraphEmbed_git_2020}.
\end{enumerate}

\begin{remark}
\label{rem:GraphKKE}
    The GraphKKE algorithm approximates transfer operators using kernel evaluations applied to snapshot adjacency matrices. However, this approach is not always suitable for comparing community structures across snapshots (see Remark \ref{rem:kernel_adj_comparison}). Even though in our work we use the Gaussian kernel and the GraphKKE implementation by Melnyk et al. \cite{Melnyk_2020} incorporates both Gaussian and Weisfeiler-Lehman kernels, we chose to focus on the latter implementation for our experiments. This decision was guided by the authors' findings, which reported superior performance of the Weisfeiler-Lehman kernel over the Gaussian kernel.
\end{remark}

{\tiny
\begin{longtable}{lc|c||c|c||c|c||c|c||c|c||c|c}
\caption[]{Adjusted Rand Index (ARI) and Normalized Mutual Information (NMI) between ground truth and snapshot clustering using different methods. An ARI or NMI value of 1 indicates perfect agreement with the ground truth, while a value of 0 indicates no agreement. Bold values indicate the highest ARI and NMI scores for each dataset. Note that the Cell Division Dataset is omitted because no universal ground truth clustering is provided; instead, different phases and subphases are identified, with their number ranging from 2 to 11. In this case, in Section \ref{sec:cell_division} we compare the results of our method with those reported in the original paper \cite{Lucas_2023} across all numbers of detected phases.}
\\
\toprule
\centering%
\textbf{Dataset} 
& \multicolumn{2}{c}{\textbf{LNE (ours)}} 
& \multicolumn{2}{c}{\textbf{IMC (ours)}} 
& \multicolumn{2}{c}{\textbf{PCA}} 
& \multicolumn{2}{c}{\textbf{GraphKKE}} 
& \multicolumn{2}{c}{\textbf{T+CL}} 
& \multicolumn{2}{c}{\textbf{tdGraphEmbed}} \\
\cmidrule(lr){2-3}
\cmidrule(lr){4-5}
\cmidrule(lr){6-7}
\cmidrule(lr){8-9}
\cmidrule(lr){10-11}
\cmidrule(lr){12-13}
& \textbf{ARI} & \textbf{NMI}
& \textbf{ARI} & \textbf{NMI}
& \textbf{ARI} & \textbf{NMI}
& \textbf{ARI} & \textbf{NMI}
& \textbf{ARI} & \textbf{NMI}
& \textbf{ARI} & \textbf{NMI} \\
\midrule
\endfirsthead

\toprule
\textbf{Dataset} 
& \multicolumn{2}{c}{\textbf{LNE (ours)}} 
& \multicolumn{2}{c}{\textbf{IMC (ours)}} 
& \multicolumn{2}{c}{\textbf{PCA}} 
& \multicolumn{2}{c}{\textbf{GraphKKE}} 
& \multicolumn{2}{c}{\textbf{T+CL}} 
& \multicolumn{2}{c}{\textbf{tdGraphEmbed}} \\
\cmidrule(lr){2-3}
\cmidrule(lr){4-5}
\cmidrule(lr){6-7}
\cmidrule(lr){8-9}
\cmidrule(lr){10-11}
\cmidrule(lr){12-13}
& \textbf{ARI} & \textbf{NMI}
& \textbf{ARI} & \textbf{NMI}
& \textbf{ARI} & \textbf{NMI}
& \textbf{ARI} & \textbf{NMI}
& \textbf{ARI} & \textbf{NMI}
& \textbf{ARI} & \textbf{NMI} \\
\midrule
\endhead

Community Split & \textbf{1} & \textbf{1} & 0.5 & 0.43 & 0.76 & 0.78 & 0.03 & 0.17 & \textbf{1} & \textbf{1} & 0 & 0.06\\
\rowcolor[gray]{0.95}
Community Hier. (2 phases) & \textbf{1} & \textbf{1} & 0.14 & 0.3 & 0.94 & 0.93 & 0.18 & 0.3 & 0.8 & 0.72 & 0.24 & 0.26\\
Community Hier. (3 phases) & \textbf{1} & \textbf{1} & 0.58 & 0.73 & \textbf{1} & \textbf{1} & 0.39 & 0.44 & 0.84 & 0.82 & 0.32 & 0.27\\
\rowcolor[gray]{0.95}
Opinion Dynamics & 0.84 & 0.77 & 0 & 0 & 0 & 0.13 & \textbf{0.88} & \textbf{0.82} & 0 & 0 & 0.32 & 0.26\\
Cholera & \textbf{0.88} & \textbf{0.84} & 0.77 & 0.73 & 0.77 & 0.73  & \textbf{0.88} & 0.83 & 0.74 & 0.7 & 0.05 & 0\\
\rowcolor[gray]{0.95}
Primary School Contacts & \textbf{0.92} & \textbf{0.9} & 0.84 & 0.81 & 0.91 & 0.89 & 0.84 & 0.82 & 0.32 & 0.43 & 0.28 & 0.32\\
\bottomrule
\label{tab:comparison}
\end{longtable}
}

The results from the comparative analysis as outlined in Table \ref{tab:comparison} emphasize the strengths and limitations of the evaluated methods, particularly the robustness and flexibility of LNE in detecting community transitions across diverse datasets. While LNE demonstrated superior performance on most datasets, its computational demands and areas for potential enhancement invite further investigation. In the following section, we reflect on these insights, discussing the broader implications of our results and outlining directions for future work to address the remaining challenges.

\section{Discussion \& Outlook}
\label{sec:discussion}

The results presented for the synthetic and real-world datasets highlight the robustness of the proposed Local Neighborhood Exploration (LNE) and Invariant Measure Comparison (IMC) methods. LNE demonstrated superior performance in detecting community transitions and identifying metastable phases, particularly in scenarios with complex dynamics, such as hierarchical structures or gradual transitions. IMC on the other hand, while computationally more efficient, was less effective in capturing subtle changes in more complex scenarios. The comparative analysis revealed that existing methods often require significant preprocessing or struggle with weighted networks, limiting their applicability. In contrast, LNE effectively uncovered transitions and stable phases in challenging datasets, such as the Cholera and Opinion Dynamics examples, showcasing its ability to provide actionable insights for real-world systems, coming from diverse domains such as biology or the social sciences.

Although LNE demonstrates high accuracy, there remain some limitations and opportunists for further development. For example, for large-scale networks, the algorithm's computational cost is an important aspect to consider. Future work will explore optimization techniques, such as sparse matrix representations or adaptive exploration times, to improve scalability of our approach. Another limitation is that our approach performs the best for clear and well-defined communities in time and space.
However, not all snapshots need to belong to exactly one phase. Similarly, as illustrated in the guiding example, not all nodes have to be members of communities (\textit{member nodes}). Random walks originating from nodes that do not belong to any community (\textit{free nodes}) will gravitate towards their closest community since they move quickly through simple regions of the network. It is important to note that our approach may not yield satisfactory results when the internal structure of communities remains stable, but a drastic reorganization among free nodes occurs. However, this can potentially be mitigated by initially detecting the member nodes and extracting the strength of connections between communities from the rest of the network. Our method can then be applied to this reduced dataset, focusing on tracking disruptions within communities. Future work will address these challenges in several ways. We plan to extend the method to account for fuzzy clustering of time snapshots, where snapshots are not rigorously attributed to one phase, but instead, they could be assigned to a certain phase with a membership value. This would allow for more flexibility when analyzing the network structure and detecting change points. In particular, this could help us to identify transition periods between phases, such as merging or splitting, during which snapshots are not assigned exclusively to only one phase. In Sections \ref{sec:syntdataexp} and \ref{sec:realdataexp}, we considered examples where all nodes are member nodes, while mixed scenarios involving both free and member nodes increase the complexity of the problem and will therefore be explored in greater detail in the future. For such cases, we plan to further reduce the computational complexity of our algorithm by differentiating between member and free nodes. As 
 shown in \cite{BKKBDS18} for stochastic differential equations, random processes starting in the same community behave similarly on intermediate timescales, the corresponding transition densities approximately form a low-dimensional manifold. This manifold can then be parametrized using manifold learning approaches. By extending this approach to the graph setting, it might be possible to efficiently extract crucial information about community structures, filtering out less important details. We expect that these improvements will enhance the scalability and overall performance of the method, making it even more effective for handling complex problems.

\section{Conclusion}
\label{sec:conclusion}
In this paper, we introduced a novel method for temporal network analysis based on random walk processes, which effectively captures community structure changes in temporal networks. The method identifies periods of network stability (network phases) and detects significant structural changes, such as splits, merges, or the formation and dissolution of communities. To this end, the proposed approach makes use of random walk processes on spatial and temporal scales to first learn the community structure of snapshots and then partition them into phases of structural stability. The method also learns low-dimensional vector representations of snapshots such that they preserve similarities in community structure and reflect each snapshot's phase affiliation within the embedding space. Furthermore, we proposed a new, highly customizable benchmark network generator for testing the efficacy of our algorithm. We further validated our method on various synthetic and real-world datasets and demonstrated that it performs better than several existing state-of-the-art algorithms. These experiments underscored its versatility and robustness across a range of applications, from social dynamics to biological systems. Therefore, our method provides a powerful tool for analyzing complex, temporal networks and gives valuable insights into the dynamics of large-scale systems.

Our approach shows promising results on various synthetic and real-world datasets. While our method has a quadratic dependence on the number of nodes per snapshot due to pairwise transition matrix comparisons, it remains computationally efficient for sparse networks and moderate numbers of snapshots, making it well-suited for practical large-scale temporal network analysis. Future work will aim to further enhance scalability by investigating optimizations such as leveraging sparse matrix representations and employing adaptive exploration times to reduce computational complexity while preserving accuracy. At the same time, our method learns low-dimensional representations of entire snapshots of temporal networks, a capability for which only a limited number of techniques exist in the current literature, as discussed in Section \ref{sec:methods_for_community_detection}. Furthermore, our method outperforms state-of-the-art representation learning methods in both community structure and phase detection accuracy. The robustness is also evident in networks with a gradual transition between phases, where other methods often struggle.

\section*{Data Availability}
The datasets generated and/or analyzed during the current study are available from the following sources: The newly generated synthetic datasets are available in the Zenodo repository titled \textit{Random walk-based snapshot clustering for detecting community dynamics in temporal networks} at \url{https://zenodo.org/records/15397630}. The Cholera dataset is available in the \textit{graphKKE} repository at \url{https://github.com/k-melnyk/graphKKE}. The Cell Division dataset is available in the \textit{phasik} repository at \url{https://gitlab.com/habermann_lab/phasik/-/tree/master?ref_type=heads}. The Primary School Contacts dataset is available at \url{http://www.sociopatterns.org/datasets/primary-school-temporal-network-data/}. The Opinion Dynamics dataset is available from the corresponding author upon reasonable request.

\section*{Code Availability}
The code used in this work is available at \cite{Blaskovic_2025}.

\bibliographystyle{abbrv} 
\bibliography{references}

\section*{Author Contributions}
T.C., N.D.C. and S.K. were involved in planning and supervised the work. F.B. and N.D.C. developed the theoretical foundation of the method. F.B. implemented the method and performed the numerical simulations and experiments. All authors discussed the results and contributed to the final manuscript.

\section*{Funding}
This work was supported by the German Ministry for Education and Research (BMBF) within the Berlin Institute for the Foundations of Learning and Data---BIFOLD (project grants 01IS18025A and 01IS18037I) and the Forschungscampus MODAL (project grant 3FO18501) and by the Deutsche Forschungsgemeinschaft (DFG, German Research Foundation) under Germany’s Excellence Strategy via MATH+: The Berlin Mathematics Research Center (EXC-2046/1, project ID: 390685689). 

\section*{Additional Information}
The authors declare no competing interests. 

\end{document}